\documentclass[sigconf]{acmart}

\renewcommand\footnotetextcopyrightpermission[1]{} %
\usepackage{algpseudocode}
\usepackage[linesnumbered,ruled,vlined]{algorithm2e}
\SetKwProg{Procedure}{Procedure}{}{}
\SetKwInOut{Input}{Input}
\SetKwInOut{Output}{Output}
\usepackage{amsfonts}
\usepackage{amsthm}
\usepackage{amsmath}

\usepackage{cleveref}
\usepackage{footnote}
\usepackage{bm}
\usepackage{subfigure}
\usepackage{framed}
\usepackage[inline]{enumitem}   
\makeatletter
\newcommand{\inlineitem}[1][]{%
	\ifnum\enit@type=\tw@
	{\descriptionlabel{#1}}
	\hspace{\labelsep}%
	\else
	\ifnum\enit@type=\z@
	\refstepcounter{\@listctr}\fi
	\quad\@itemlabel\hspace{\labelsep}%
	\fi}

\makesavenoteenv{tabular}
\makesavenoteenv{algorithm}

\newcommand{\junk}[1]{{}}
\newcommand{\SC}{\mathcal{NSC}}
\newcommand{\USC}{\mathcal{USC}}
\newcommand{\sweep}{\mathrm{sweep}}
\newcommand{\appkmeans}{\Call{$k$-means}{}}

\newcommand{\fail}{\mathrm{fail}}
\newcommand{\APP}{\mathrm{APP}}

\newcommand{\E}{\mathop{\mathbf{E}}}

\newcommand{\supp}{\mathrm{supp}}

\newcommand{\Alg}{\mathcal{A}}

\newcommand{\vol}{\mathrm{vol}}

\newcommand{\1}{\mathbf{1}}

\newcommand{\pnote}[1]{\textcolor{blue}{(Pan: #1)}}

\newtheorem{theorem}{Theorem}[section]

\newtheorem{lemma}[theorem]{Lemma}

\newtheorem{claim}[theorem]{Claim}

\theoremstyle{definition}

\newtheorem{assumption}[theorem]{Assumption}

\providecommand{\abs}[1]{\left\lvert#1\right\rvert}
\providecommand{\norm}[1]{\left\lVert#1\right\rVert}

\allowdisplaybreaks
\begin{document}
	\fancyhead{}

\title{Average Sensitivity of Spectral Clustering}
\author{Pan Peng}
\affiliation{%
	\institution{Department of Computer Science\\University of Sheffield}
}
\email{p.peng@sheffield.ac.uk}
\orcid{0000-0003-2700-5699}

\author{Yuichi Yoshida}
\orcid{}
\affiliation{%
	\institution{National Institute of Informatics}
  \institution{JST, PRESTO}
}
\email{yyoshida@nii.ac.jp}

\begin{abstract}
  Spectral clustering is one of the most popular clustering methods for finding clusters in a graph, which has found many applications in data mining.
  However, the input graph in those applications may have many missing edges due to error in measurement, withholding for a privacy reason, or arbitrariness in data conversion.
  To make reliable and efficient decisions based on spectral clustering, we assess the stability of spectral clustering against edge perturbations in the input graph using the notion of average sensitivity, which is the expected size of the symmetric difference of the output clusters before and after we randomly remove edges.

  We first prove that the average sensitivity of spectral clustering is proportional to $\lambda_2/\lambda_3^2$, where $\lambda_i$ is the $i$-th smallest eigenvalue of the (normalized) Laplacian.
  We also prove an analogous bound for $k$-way spectral clustering, which partitions the graph into $k$ clusters.
  Then, we empirically confirm our theoretical bounds by conducting experiments on synthetic and real networks.
  Our results suggest that spectral clustering is stable against edge perturbations when there is a cluster structure in the input graph.
\end{abstract}



\maketitle

\section{Introduction}

Spectral clustering is one of the most popular graph clustering methods, which finds tightly connected vertex sets, or \emph{clusters}, in the input graph using the eigenvectors of the associated matrix called the (normalized) Laplacian.
It has been used in many applications such as image segmentation~\cite{Shi2000}, community detection in networks~\cite{Fortunato2010}, and manifold learning~\cite{Belkin2001}.
See~\cite{Luxburg2007} for a survey on the theoretical background and practical use of spectral clustering.

In those applications, however, the input graph is often untrustworthy, and the decision based on the result of spectral clustering may be unreliable and inefficient.
We provide some examples below.
\begin{itemize}
\item A social network is a graph whose vertices correspond to users in a social network service (SNS), and two vertices are connected if the corresponding users have a friendship relation. However, users may not report their friendship relations because they do not actively use the SNS, or they want to keep the relationship private.
\item A sensor network is a graph whose vertices correspond to sensors allocated in some space, and two vertices are connected if the corresponding sensors can communicate with each other.
The obtained sensor network may be untrustworthy because some sensors might be unable to communicate due to power shortage or obstacles temporarily put between them.
\item In manifold learning, given a set of vectors, we construct a graph whose vertices correspond to the vectors, and we connect two vertices if the corresponding vectors have a distance below a certain threshold. The choice of the threshold is arbitrary, and the obtained graph can vary with the threshold.
\end{itemize}
If the output clusters are sensitive to edge perturbations, we may make a wrong decision or incur some cost to cancel or update the decision.
Hence, to make a reliable and efficient decision using spectral clustering under these situations, the output clusters should be stable against edge perturbations.

One might be tempted to measure the size of the symmetric difference of the output clustering before and after \emph{adversarial} edge perturbations.
However, spectral clustering is sensitive to adversarial edge perturbations.
For example, suppose that we have a connected graph $G$ with two ``significant'' clusters $S,\overline{S}$, i.e., the subgraphs induced by $S$ and $\overline{S}$ are well-connected inside while there are few edges between $S$ and $\overline{S}$. It is known that the spectral clustering will output a set, which, together with its complement, corresponds to a bipartition that is close to the partition $\{S,\overline{S}\}$~\cite{KLLOT:improved}. However, deleting all the edges incident to any vertex $v$ will result in a new graph $G'$ on which the spectral clustering will output $v$, or equivalently, the partition $\{\{v\}, \overline{\{v\}}\}$. That is, the output clustering in the original graph is very different from the one in the perturbed graph.

In general, spectral clustering is sensitive to the noisy ``dangling sets'', which are connected to the core of the graph by only one edge~\cite{zhang2018understanding}. This suggests that the above way of measuring the stability of spectral clustering might be too pessimistic.

\subsection{Our contributions}
In this work, we initiate a systematic study of the stability of spectral clustering against edge perturbations, using the notion of average sensitivity~\cite{VY19:sensitivity}, which is the expected size of the symmetric difference of the output clusters before and after we \emph{randomly} remove a few edges. Using average sensitivity is more appropriate in many applications, in which the aforementioned adversarial edge perturbations rarely occur. Furthermore, if we can show that the average sensitivity is at most $\beta$, then by Markov's inequality, for the 99\% of possible edge perturbations, the symmetric difference size of the output cluster is bounded by $100\beta$, which further motivates the use of average sensitivity.

We first consider the simplest case of partitioning a graph into two clusters: the algorithm computes the eigenvector corresponding to the second smallest eigenvalue of Laplacian and then outputs a set according to a sweep process over the eigenvector.
For both unnormalized and normalized Laplacians, we show that if the input graph has a ``significant'' cluster structure, then the average sensitivity of spectral clustering is proportional to $\lambda_2/\lambda_3^2$, where $\lambda_i$ is the $i$-th smallest eigenvalue of the corresponding Laplacian.

This result is intuitively true because if $\lambda_2/\lambda_3^2$ is small, then by higher-order Cheeger's inequality~\cite{KLLOT:improved}, the graph can be partitioned into two intra-dense clusters with a few edges between them. That is, the cluster structure is significant.
Hence, we are likely to output the same cluster after we randomly remove a few edges.

\begin{figure}[t!]
  \subfigure[]{\includegraphics[width=.3\hsize]{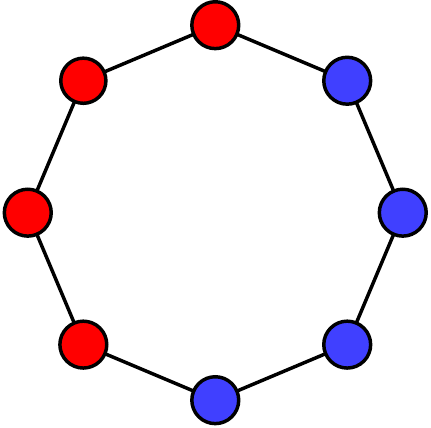}}
  \quad
  \subfigure[]{\includegraphics[width=.3\hsize]{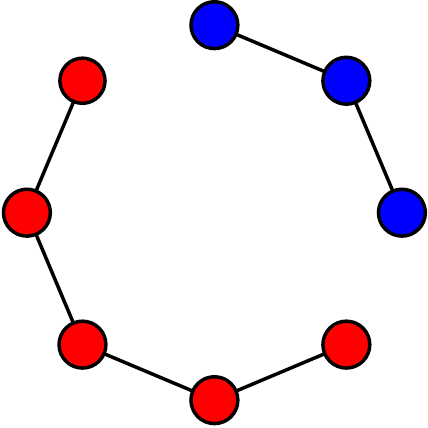}}
  \quad
  \subfigure[]{\includegraphics[width=.3\hsize]{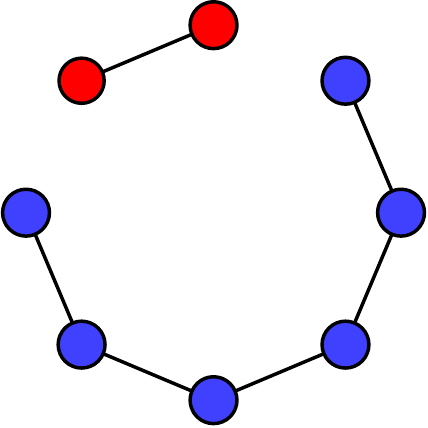}}
  \caption{Cycle graph with 8 vertices and the result of 2-way clustering. The output clusters are drastically different between (a) the original graph and (b), (c) the graphs with some edges removed.}\label{fig:cycle}
\end{figure}

In contrast, a typical graph with a high $\lambda_2/\lambda_3^2$ is an $n$-cycle (Figure~\ref{fig:cycle}).
We can observe that the spectral clustering is not stable on this graph because we get drastically different connected components depending on the choice of removed edges.

Next, we consider $k$-way spectral clustering: the algorithm computes the first $k$ eigenvectors of Laplacian, and then outputs a clustering by invoking the $k$-means algorithm on the embedding corresponding to the eigenvectors.
We consider the spectral clustering algorithm that exploits the normalized Laplacian, which was popularized by Shi and Malik~\cite{Shi2000}. We show that the average sensitivity of the algorithm is proportional to $\sqrt{\lambda_k}/\lambda_{k+1}$, which again matches the intuition. %

Finally, using synthetic and real networks, we empirically confirm that the average sensitivity of spectral clustering correlates well with $\lambda_2/\lambda_3^2$ or $\sqrt{\lambda_k}/\lambda_k$, and that it grows linearly in the edge removal probability, as our theoretical results indicate.

Our theoretical and empirical findings can be summarized as follows.
\begin{framed}
  We can reliably use spectral clustering: It is stable against random edge perturbations if there is a significant cluster structure in a graph, and it is irrelevant otherwise.
\end{framed}

\subsection{Organization}
We discuss related work in Section~\ref{sec:related}, and we introduce notions that will be used throughout this paper in Section~\ref{sec:pre}.
Then, we study the average sensitivity of $2$-way spectral clustering with unnormalized and normalized Laplacians in Sections~\ref{sec:unnormalized} and~\ref{sec:normalized}, respectively.
We discuss the average sensitivity of $k$-way spectral clustering in Section~\ref{sec:kcluster}.
We provide our empirical results in Section~\ref{sec:experiments} and conclude our work in Section~\ref{sec:conclusions}.

\section{Related Work}\label{sec:related}

Huang~et~al.~\cite{Huang2008} considered manifold learning, in which we construct a graph $G$ on $n$ vertices from a set of $n$ vectors $\bm{v}_1,\ldots,\bm{v}_n$ by connecting the $i$-th vertex and the $j$-th vertex by an edge with weight determined by the distance $\|\bm{v}_i-\bm{v}_j\|$, and then apply spectral clustering on $G$.
They analyzed how the (weighted) adjacency matrix of $G$, the normalized Laplacian $L$ of $G$, and the second eigenvalue/eigenvector of $L$ change when we perturb the vectors $\bm{v}_1,\ldots,\bm{v}_n$.
Our work is different in two points.
Firstly, we consider random edge perturbations rather than perturbation to the vector data.
Hence, we can apply our framework to more general contexts, such as social networks and sensor networks.
Secondly, we directly analyze the output clusters rather than the intermediate eigenvalues and eigenvectors.

Bilu and Linial introduced the notion of \emph{stable instances} for clustering problems to model realistic instances with an outstanding solution that is stable under noise~\cite{BL12:stable}.
In their setting, for a clustering problem (e.g., sparsest cut) and a parameter $\alpha\geq 1$, a graph with edge or vertex weights is said to be \emph{$\alpha$-stable} if the optimal solution does not change when we perturb all the edge/vertex weights by a factor of at most $\alpha$.
It has been shown that some clustering problems are solvable in polynomial time on stable instances. Karrer~et~al.~\cite{karrer2008robustness} considered the robustness of the community (or cluster) structure of a network by perturbing it as follows: Remove each edge with probability $p$, and replace it with another edge between a pair of vertices $(i,j)$, chosen at random with an appropriate probability. They used modularity-based methods to find clusters of the original unperturbed graph and the perturbed one, and measured the variation of information between the corresponding two partitions. Gfeller~et~al.~\cite{gfeller2005finding} considered the robustness of the cluster structure of weighted graphs where the perturbation is introduced by randomly increasing or decreasing the edge weights by a small factor. In contrast to modeling stable instances for clustering~\cite{BL12:stable} or studying the stability of a partition against random perturbations~\cite{karrer2008robustness,gfeller2005finding}, our study focuses on the stability of the spectral clustering algorithms.

Our work is also related to a line of research on eigenvalue perturbation, which studies how close (or how far) the eigenvalues of a matrix $M+H$ to those of $M$, where $H$ is ``small'' in some sense and $M+H$ is viewed as a perturbation of $M$~\cite{stewart1990matrix}.
The classical theorem due to Weyl (see Theorem~\ref{thm:weyl}) bounds, for each $i\leq n$, the differences between the $i$-th eigenvalue of $M+H$ and that of $M$ by the spectral norm of $H$.
Eldridge~et~al.~recently gave some perturbation bounds on the eigenvalues and eigenvectors when the perturbation $H$ is random~\cite{eldridge2018unperturbed}. There also exist studies on the eigenvalues (and eigenvectors) of $M(\varepsilon)$ that is a matrix function of a parameter $\varepsilon$ and is analytic in a small neighborhood of some value $\varepsilon_0$, satisfying $M(\varepsilon_0)=M$ (e.g.,~\cite{kato2013perturbation}). However, we cannot apply such results to our setting, as even for a single edge deletion, we need to consider beyond the small neighborhood.

\section{Preliminaries}\label{sec:pre}

Let $\bm{1} \in \mathbb{R}^n$ be the all-one vector.
When all the eigenvalues of a matrix $A \in \mathbb{R}^{n \times n}$ are real, we denote by $\lambda_i(A)$ the $i$-th smallest eigenvalue of $A$.
Also, we denote by $\lambda_{\min}(A)$ and $\lambda_{\max}(A)$ the smallest and largest eigenvalues of $A$, respectively.

We often use the symbols $n$, $m$, and $\Delta$ to denote the number of vertices, the number of edges, and the maximum degree, respectively, of the graph we are concerned with, which should be clear from the context.
For a graph $G=(V,E)$ and a vertex set $S \subseteq V$, $G[S]$ denotes the subgraph of $G$ induced by $S$.
The \emph{volume} $\mathrm{vol}(S)$ of $S$ is the sum of degrees of vertices in $S$.

\subsection{Average Sensitivity}

In order to measure the sensitivity of spectral clustering algorithms, which partition the vertex set of a graph into $k$ clusters for $k \geq 2$, we adapt the original definition of average sensitivity~\cite{VY19:sensitivity} as follows.

Let $G=(V,E)$ be a graph.
For an edge set $F \subseteq E$, we denote by $G-F$ the graph $(V,E \setminus F)$.
For $p \in [0,1]$, we mean by $F \sim_p E$ that each edge in $E$ is included in $F$ with probability $p$ independently from other edges.

For vertex sets $S,T \subseteq V$, Let $S\triangle T$ denote the symmetric difference between two vertex sets $S,T$.
Let $\mathcal{P}=\{P_1,\ldots, P_k\}$ and $\mathcal{Q}=\{Q_1,\ldots,Q_k\}$ be two $k$-partitions of $V$.
Then, the \emph{distance} of $\mathcal{P}$ and $\mathcal{Q}$ with respect to vertex size (resp., volume) is defined as follows:
\begin{align*}
  d_{\mathrm{size}}(\mathcal{P},\mathcal{Q}) & =\min_{\sigma}\sum_{i=1}^k |P_i\triangle Q_{\sigma(i)}|, \\
  (\text{resp., }d_{\mathrm{vol}}(\mathcal{P},\mathcal{Q}) & = \min_{\sigma}\sum_{i=1}^k \vol(P_i\triangle Q_{\sigma(i)})),
\end{align*}
where $\sigma$ ranges over all bijections $\sigma\colon \{1,\ldots,k\}\rightarrow \{1,\ldots,k\}$.
It is easy to see that $d_{\mathrm{size}}$ and $d_{\mathrm{vol}}$ satisfy the triangle inequality.
When $k=2$ and $\mathcal{P} = \{P,\overline{P}\}$ and $\mathcal{Q} = \{Q,\overline{Q}\}$, we simply write $d_\mathrm{size}(P,Q)$ and $d_{\mathrm{vol}}(P,Q)$ instead of $d_\mathrm{size}(\mathcal{P},\mathcal{Q})$ and $d_{\mathrm{vol}}(\mathcal{P},\mathcal{Q})$.

For an algorithm $\Alg$ that outputs a $k$-partition and a real-valued function $\beta$ on graphs, we say that the \emph{$p$-average sensitivity} with respect to vertex size (resp.,  volume) of $\Alg$ is at most $\beta$ if
\begin{align*}
  \E_{F\sim_p E}[d_{\mathrm{size}}(\Alg(G),\Alg(G-F))] & \leq \beta(G), \\
  (\text{resp., }\E_{F\sim_p E}[d_{\mathrm{vol}}(\Alg(G),\Alg(G-F))] & \leq \beta(G)).
\end{align*}
We note that this definition is different from the original one~\cite{VY19:sensitivity} in that we remove each edge with probability $p$ independently from others whereas the original one removes $k$ edges without replacement for a given integer parameter $k$.

\subsection{Spectral Graph Theory}

\subsubsection{Notions from spectral graph theory}
Let $G=(V,E)$ be an $n$-vertex graph.
We always assume that the vertices are indexed by integers, i.e., $V = \{1,\ldots,n\}$.
For a vertex set $S$, we denote by $\overline{S}$ the complement set $V \setminus S$.%

For two vertex sets $S,T\subseteq V$, $E(S,T)$ denotes the set of edges between $S$ and $T$, that is, $E(S,T) = \{(i,j) \in E : i \in S, j \in T\}$.
The \emph{cut ratio} of $S \subseteq V$ is defined to be $\alpha_G(S)=\frac{|E(S,\overline{S})|}{\min\{|S|,|\overline{S}|\}}$, and the \emph{cut ratio} of $G$ is defined to be $\alpha(G)=\min_{\emptyset \subsetneq S \subsetneq V}\alpha_G(S)$.
The \emph{conductance} of a set $S$ is defined to be $\phi_G(S)=\frac{|E(S,\overline{S})|}{\min\{\vol(S),\vol(\overline{S})\}}$, and the \emph{conductance} of $G$ is defined to be $\phi(G)=\min_{\emptyset \subsetneq S \subsetneq V}\phi_G(S)$.
For an integer $k$, let $\rho_G(k)$ be the \emph{$k$-way expansion} defined as
\[
\rho_G(k) := \min_{S_1,\ldots,S_k:\text{ partition of }V}\max_{1\leq i\leq k}\phi_G(S_i).
\]

The \emph{adjacency matrix} $A_G \in {\{0,1\}}^{n \times n}$ of $G$ is defined as ${(A_G)}_{ij} = 1$ if and only if $(i,j) \in E$.
The \emph{degree matrix} $D_G \in \mathbb{R}^{n \times n}$ of $G$ is the diagonal matrix with ${(D_G)}_{ii} = d_i$, where $d_i$ is the degree of the $i$-th vertex.
The \emph{Laplacian} $L_G \in \mathbb{R}^{n \times n}$ of $G$ is defined as $L_G=D_G-A_G$.
The \emph{normalized Laplacian} $\mathcal{L}_G \in \mathbb{R}^{n\times n}$ of $G$ is defined as $\mathcal{L}_G = D_G^{-1}L_G  =I-D_G^{-1}A_G$.\footnote{In some literature, $\mathcal{L}$ is called the \emph{random-walk Laplacian}.}
It is well known that all the eigenvalue of $L_G$ and $\mathcal{L}_G$ are nonnegative real numbers.
We sometimes call $L_G$ the \emph{unnormalized Laplacian} of $G$.

We omit subscripts if they are clear from the context.

\subsubsection{Cheeger's inequality}

\begin{algorithm}[t!]
\caption{Sweep algorithms}\label{alg:sweep}
\Procedure{$\sweep_\alpha(\bm{v})$}{
	Reorder the vertices in $G$ in non-decreasing order in terms of $\bm{v}$, i.e., $v_1 \leq v_2 \leq \cdots\leq v_n$\;
	\Return the set with the minimum \emph{cut ratio} among the sets of the form $\{1,\ldots,i\}\;(1 \leq i\leq n-1)$.
}
\Procedure{$\sweep_\phi(\bm{v})$}{
	Reorder the vertices in $G$ in non-decreasing order in terms of $\bm{v}$, i.e., $v_1 \leq v_2 \leq \cdots\leq v_n$\;
	\Return the set with the minimum \emph{conductance} among the sets of the form $\{1,\ldots,i\}\;(1 \leq i\leq n-1)$.
}
\end{algorithm}

We can use unnormalized and normalized Laplacians to find vertex sets with small cut ratio and conductance, respectively.
Consider procedures in Algorithm~\ref{alg:sweep}, which keep adding vertices in the order determined by the given vector $\bm{v}$, and then return the set with the best cut ratio and conductance.
Then, the following inequality is known.
\begin{lemma}[Cheeger's inequality~\cite{Alon:1986gz,Alon:1985jg}]\label{lem:Cheeger}
  We have
  \[
    \frac{\lambda_2}{2}\leq \alpha(G)\leq \sqrt{2\Delta\lambda_2}
    \quad \text{and} \quad
    \frac{\nu_2}{2}\leq \phi(G)\leq \sqrt{2\nu_2},
  \]
  where $\lambda_2 = \lambda_2(L_G)$ and $\nu_2 = \lambda_2(\mathcal{L}_G)$.
  In particular, $\alpha(\sweep_\alpha(\bm{v}_2)) \leq \sqrt{2\Delta\lambda_2}$ and $\phi(\sweep_\phi(\bm{\bar{v}}_2)) \leq \sqrt{2\nu_2}$ hold, where $\bm{v}_2$ and $\bm{\bar{v}}_2$ are the %
  eigenvectors of $L_G$ and $\mathcal{L}_G$ corresponding to $\lambda_2$ and $\nu_2$, respectively.
\end{lemma}

The following variant of Cheeger's inequality is also known.
\begin{lemma}[Improved Cheeger's inequality~\cite{KLLOT:improved}]\label{lem:improvedCheeger}
	For any $k\geq 2$, we have
	\[
	  \alpha(\sweep_\alpha(\bm{v}_2))= O\left (\frac{k\lambda_2\sqrt{\Delta}}{\sqrt{\lambda_k}}\right),
  	\text{ and }
  	\phi(\sweep_\phi(\bm{\bar{v}}_2))= O\left (\frac{k\nu_2}{\sqrt{\nu_k}}\right),
	\]
	where $\lambda_k = \lambda_k(L_G)$, $\nu_k=\lambda_k(\mathcal{L}_G)$, and $\bm{v}_2$ and $\bm{\bar{v}}_2$ are the (right) eigenvectors of $L_G$ and $\mathcal{L}_G$ corresponding to $\lambda_2$ and $\nu_2$, respectively.
\end{lemma}
As $\lambda_k \geq \lambda_2$ and $\nu_k \geq \nu_2$, the bounds in Lemma~\ref{lem:improvedCheeger} are always stronger than those in Lemma~\ref{lem:Cheeger}.
We note that the original proof of~\cite{KLLOT:arxiv} only handled the normalized case.
However, their proof can be easily adapted to the unnormalized case, which we discuss in Appendix \ref{sec:proof-of-improved-cheeger}.

For $k$-way expansion, the following higher-order Cheeger inequality is known.
\begin{theorem}[\cite{lee2014multiway}]\label{thm:higherCheeger}
  We have
  \[
    \frac{\mathbf{\lambda}_k(\mathcal{L}_G)}{2}\leq \rho_G(k)\leq O\left (k^2\sqrt{\lambda_k(\mathcal{L}_G)}\right).
  \]
\end{theorem}

\subsubsection{Spectral Clustering}\label{subsubsec:spectral-clustering}

\begin{algorithm}[t!]
  \caption{Spectral Clustering}\label{alg:spectral-clustering}
  \Procedure{$\USC_2(G)$}{
    Compute the second eigenvector $\bm{v}_2$ of $L_G$\;
    \Return $\sweep_\alpha(\bm{v}_2)$.
  }
  \Procedure{$\SC_{2}(G)$}{
    Compute the second eigenvector $\bm{v}_2$ of $\mathcal{L}$\;
    \Return $\sweep_\phi(\bm{v}_2)$.
  }
\end{algorithm}

In this work, we consider the spectral clustering algorithms described in Algorithm~\ref{alg:spectral-clustering}.
In the unnormalized case ($\USC_2$), we first compute the second eigenvector $\bm{v}_2 \in \mathbb{R}^n$ of the Laplacian of the input graph and then return the set computed by running the $\sweep_\alpha$ procedure on $\bm{v}_2$.
In the normalized case ($\SC_2$), we replace $\bm{v}_2$ with the second eigenvector of the normalized Laplacian and give it to the $\sweep_\phi$ procedure.

\begin{algorithm}[t!]
  \caption{Spectral Clustering with $k$ clusters}\label{alg:kcluster}
  \Procedure{$\SC_{k\mathrm{-means}}(G)$}{
    Compute the first $k$ eigenvectors $\bm{v}_1,\ldots,\bm{v}_k$ of $\mathcal{L}_G$\;
    Embedding each vertex $u \in V$ to $f(u)=(\bm{v}_1(u),\ldots, \bm{v}_k(u))$\;
    \Return \Call{$k$-means}{$f(u_1),\ldots,f(u_n)$}.
  }
\end{algorithm}

Algorithm~\ref{alg:kcluster} is a variant of Algorithm~\ref{alg:spectral-clustering} that partitions the graph into $k$ clusters.
Here, we first compute the top $k$ eigenvectors $\bm{v}_1,\ldots,\bm{v}_k$, and embed each vertex $u \in V$ to a point $(\bm{v}_1(u),\ldots,\bm{v}_k(u))$ in  $\mathbb{R}^k$.
Then, we apply the $k$-means algorithm~\cite{MacQueen1967} to obtain $k$ clusters.
This algorithm ($\SC_{k\mathrm{-means}}$), that makes use of  $\mathcal{L}_G=I-D_G^{-1}A_G$, was popularized by Shi and Malik~\cite{Shi2000}. There are also other versions of spectral clustering based on the $k$-means algorithm (see~\cite{Luxburg2007}). %
We remark that the one we are analyzing is preferred in practice. For example, in the survey~\cite{Luxburg2007}, it is said ``\textit{In our opinion, there are several arguments which advocate for using normalized rather than unnormalized spectral clustering, and in the normalized case to use the eigenvectors of $L_{\textrm{rw}}$ \emph{(i.e., $\mathcal{L}_G$ as we consider here)} rather than those of $L_{\textrm{sym}}$ \emph{(i.e., $I-D_G^{-1/2}A_G D_G^{-1/2}$)}.}''

The following bound is known for $\SC_{k\text{-means}}$.
\begin{theorem}[\cite{peng2017partitioning}, rephrased]\label{thm:goodkclustering}
  Let $G$ be a graph with $\frac{\lambda_{k+1}(\mathcal{L}_G)}{\rho_G(k)}=\Omega(k^3)$.
  Let $\{S_1^*,\ldots,S_k^*\}$ be a $k$-partition of $G$ achieving $\rho_G(k)$ and let $\{A_1,\ldots, A_k \}$ be the output of $\SC_{k\textrm{-means}}$.
  If the approximation ratio of $\appkmeans$ (in terms of the objective function of $\appkmeans$) is $\alpha$, then %
  we have
\[
  d_{\textrm{vol}}(\{S_1^*,\ldots,S_k^*\},\{A_1,\ldots, A_k \})=O\left(\frac{\alpha k^3 \rho_G(k)}{\lambda_{k+1}(\mathcal{L}_G)} \cdot \vol(G)\right).
\]
\end{theorem}
We remark that in~\cite{peng2017partitioning}, the above theorem was stated in terms of the spectral clustering algorithm that uses the eigenvectors of $L_{\textrm{sym}}:=I-D^{-1/2}AD^{-1/2}$, which turns out to be equivalent to $\SC_{k\text{-means}}$.
\subsection{Stable Instances}\label{subsec:stable-instances}
We introduce the notion of stable instances, which is another tool we need to analyze average sensitivity of spectral clustering.

For $\varepsilon \in (0,1)$, and two sets $S,T\subseteq V$, we call the corresponding bipartitions $\mathcal{S}=\{S,\overline{S}\}$ and $\mathcal{T}=\{T,\overline{T}\}$ are \emph{$\varepsilon$-close} with respect to size (resp., volume) if
\[
  d_{\mathrm{size}}(\mathcal{S}, \mathcal{T}) \leq \varepsilon n \quad (\text{resp., } d_{\mathrm{vol}}(\mathcal{S}, \mathcal{T}) \leq \varepsilon \cdot \mathrm{vol}(G)).
\]
We call a graph $G=(V,E)$ \emph{$(\rho,\varepsilon)$-stable} with respect to cut ratio (resp., conductance) if any $\rho$-approximate solution $S \subseteq V$, that is, $\alpha_G(S) \leq \rho \cdot \alpha(G)$ (resp., $\phi_G(S) \leq \rho \cdot \phi(G)$), is $\varepsilon$-close to any optimal solution with respect to size (resp., volume).
The following is known.
\begin{lemma}[Corollary 4.17 in~\cite{KLLOT:arxiv}]\label{lemma:stable_kllot_Delta}
  Let $G=(V,E)$ be a graph.
  For any $\rho \geq 1$, $G$ is
  \begin{align*}
  & \left(\rho,\Theta\left( \frac{\rho \lambda_2(L_G)\Delta^{1/2}}{{\lambda_3(L_G)}^{3/2}}\right)\right)\text{-stable with respect to cut ratio, and}\\
	& \left(\rho,\Theta\left(\frac{\rho \lambda_2(\mathcal{L}_G)}{{\lambda_3(\mathcal{L}_G)}^{3/2}}\right)\right)\text{-stable with respect to conductance}.
  \end{align*}
\end{lemma}
Although Kwok~et~al.~\cite{KLLOT:arxiv} showed Lemma~\ref{lemma:stable_kllot_Delta} only for the normalized case, we can easily modify the proof for the unnormalized case, which we provide in Appendix \ref{sec:proof-of-stable_kllot_Delta}.

\subsection{Tools from Matrix Analysis} %
We will make use of the following results.

\begin{theorem}[Weyl's inequality]\label{thm:weyl}
	Let $A, H \in \mathbb{R}^{n \times n}$ be symmetric matrices.
	Let $\{\lambda_i\},\{\lambda_i'\}$ be the eigenvalues of $A$ and $A+H$, respectively. Then for any $1 \leq i\leq n$, we have
	\[
		\abs{\lambda_i-\lambda'_i}\leq \norm{H},
	\]
  where $\norm{H}$ is the spectral norm of $H$.
\end{theorem}

\begin{theorem}[Theorem 5.1.1 in~\cite{tropp2015introduction}]\label{thm:matrix_chernoff}
	Let $X=\sum_{i=1}^T X_i$, where $X_i\in \mathbb{R}^{n\times n}\;(1\leq i\leq T)$ are independent random symmetric matrices. Assume that $0\leq \lambda_{\min}(X_i)$ and $\lambda_{\max}(X_i)\leq R$ for any $1 \leq i\leq T$. Let %
	$\mu_{\max}=\lambda_{\max}(\E[X])$. Then for any $\varepsilon>0$,
	\begin{eqnarray*}
		\Pr[\lambda_{\max}(X)\geq (1+\varepsilon)\mu_{\max}]\leq n{\left(\frac{e^\varepsilon}{{(1+\varepsilon)}^{1+\varepsilon}}\right)}^{\mu_{\max}/R}
	\end{eqnarray*}
\end{theorem}

\section{Spectral Clustering with Unnormalized Laplacian}\label{sec:unnormalized}

In this section, we analyze the $p$-average sensitivity of $\USC_2$ for $p \in [0,1]$.
For a graph $G$, let $\lambda_i(G)$ denote the $i$-th smallest eigenvalue of the Laplacian associated with a graph $G$, that is, $\lambda_i(G) = \lambda_i(L_G)$.
The goal of this section is to show the following under Assumption~\ref{assume:unnormalized}, which we will explain in Section~\ref{subsec:assumption-unnormalized}.
\begin{theorem}\label{thm:spectraclustering}
  Let $G=(V,E)$ be a graph and $p \in [0,1]$.
  If Assumption~\ref{assume:unnormalized} holds, then the $p$-average sensitivity of $\USC_2$ is
  \[
    O\left(\frac{\lambda_2(G)}{{\lambda_3(G)}^2} \cdot \Delta n+1\right).
  \]
\end{theorem}

We discuss Assumption~\ref{assume:unnormalized} and its plausibility in Section~\ref{subsec:assumption-unnormalized}.
Before proving Theorem~\ref{thm:spectraclustering}, we first discuss the sensitivity of the eigenvalues in Section~\ref{subsec:eigenvalue-unnormalized}.
Then, we prove Theorem~\ref{thm:spectraclustering} in Section~\ref{subsec:spectral-clustering-unnormalized}.

\subsection{Assumptions}\label{subsec:assumption-unnormalized}

Given a graph $G$ and a value $p\in [0,p]$, the \emph{reliability} $C(p)$ of $G$ is the probability that if each edge fails with probability $p$, no connected component of $G$ is disconnected as a result \cite{colbourn1987combinatorics}.
There exists a fully polynomial-time randomized approximation scheme for $C(p)$~\cite{GJ19:polynomial}.
We will derive a bound on $p$-average sensitivity under the following assumption. Let $p_{\fail}:=O(\max\{\lambda_2(G)/{\lambda_3(G)}^2,n^{-1}\})$.
\begin{assumption}\label{assume:unnormalized}
We assume the following properties hold.
  \begin{enumerate}[label=(\roman*)]
\item[]  \inlineitem $\lambda_3(G)\geq %
  \max\{24p\Delta,48\log n \}$; 
  \inlineitem $p_{\fail}\leq 1$; 
\item[]  \inlineitem $C(p)\geq 1-{p_{\fail}}$.
  \end{enumerate}
\end{assumption}

\paragraph{Plausibility of Assumptions} Although the conditions in Assumptions~\ref{assume:unnormalized} are technical, they conform to our intuitions about graphs with low average sensitivity. A graph satisfying those conditions is naturally composed of two vertex disjoint intra-dense subgraphs $S^*$ and $\overline{S^*}$, with no or few crossing edges between them. More generally,
\begin{itemize}
\item Assumption~\ref{assume:unnormalized}(i) and (ii) imply that $\lambda_2$ is small but $\lambda_3$ is large, which imply that the graph has at most one outstanding sparse cut by the higher-order Cheeger inequality~\cite{lee2014multiway,KLLOT:improved}. It has been discussed that the (normal vectors of the) eigenspaces of Laplacian with such a large eigengap are stable against edge perturbations~\cite{Luxburg2007}. To better understand Assumption~\ref{assume:unnormalized}(i), let us consider an example. Suppose that $G$ can be partitioned into two clusters $S$ and $\overline{S}$ such that $|S|=\Theta(|\overline{S}|)$, and the induced subgraphs $G[S]$ and $G[\overline{S}]$ have conductance at least $\Omega(1)$ (i.e., the cluster structure of $G$ is significant), and the degree of each vertex in both subgraphs is in $[\Delta/4, \Delta]$. Then it holds that $\lambda_2(G[S]), \lambda_2(G[\overline{S}])=\Omega(\Delta)$ (see e.g. \cite{hoory2006expander}). This further implies that $\lambda_3(G)=\Omega(\Delta)$ \cite{lee2014multiway}, and thus $G$ satisfies the assumption as long as $\Delta =\Omega(\log n)$.
\item Assumption~\ref{assume:unnormalized} (iii) corresponds to the intuition that each connected component of the graph remains connected after removing a small set of random edges with high probability. If this is not the case, then intuitively the graph contains many ``dangling sets'' that are loosely connected to the core of the graph, in which case the algorithm is not stable~\cite{zhang2018understanding}.
\end{itemize}

Note that if the graph $G$ satisfying Assumption~\ref{assume:unnormalized} is not connected, i.e., $\lambda_2(G)=0$, then (ii) is  trivially satisfied. Then essentially the conditions become that the graph has large $\lambda_3(G)$, and thus has two connected components $S^*,\overline{S^*}$ and $p$ is reasonably small that the corresponding perturbation will not disconnect $G[S^*]$ or $G[\overline{S^*}]$ with high probability.
\subsection{Average Sensitivity of Eigenvalues of \texorpdfstring{$L_G$}{L}}\label{subsec:eigenvalue-unnormalized}

The goal of this section is to show the following.
\begin{lemma}\label{lem:eigenvalue-unnormalized}
  Let $G=(V,E)$ be a graph and $p \in [0,1]$, and let $F \sim_p E$.
  If Assumption~\ref{assume:unnormalized}(i) holds,
  then
  \[
    \lambda_3(G-F)\geq \frac{\lambda_3(G)}{2}
  \]
  holds with probability at least $1-n^{-10}$.
\end{lemma}

We define $E_{ab}\in \{0,1\}^{n \times n}$ as the matrix such that $(E_{ab})_{cd}=1$ if and only if $c=a$ and $d=b$. For each edge $e=(i,j)\in E$, we let $E_e=E_{ii}+E_{jj}-E_{ij}-E_{ij}$.  For a set $F \subseteq E$ of edges, let $E_F=\sum_{e\in F} E_e$, that is, $E_F$ is the Laplacian matrix of the graph with vertex set $V$ and edge set $F$.
Note that
\begin{align*}
  & L_{G-F}=D_{G-F}-A_{G-F} \\
  & =D-\sum_{(i,j)\in F}(E_{ii}+E_{jj})-\left(A-\sum_{(i,j)\in F}(E_{ij}+E_{ji})\right)=L_G- E_F.
\end{align*}

The following directly follows from Theorem~5 in~\cite{eldridge2018unperturbed}.
\begin{lemma}\label{lem:eigenvalue_multiple}
  Let $G=(V,E)$ be a graph, $F \subseteq E$, and $1\leq t\leq n$.
  Let $h$ be such that $\bm{x}^\top E_F \bm{x} \leq h$ for any \emph{unit} vector $\bm{x} \in \mathbb{R}^n$ in $V':=\mathrm{span}(\bm{v}_t,\ldots,\bm{v}_n)$, where $\bm{v}_i$ is the eigenvector corresponding to $\lambda_i(G)$.
  Then, we have $\lambda_t(G-F)\geq \lambda_t(G) - h$.
\end{lemma}
Next, we prove the following.
\begin{lemma}\label{lemma:random_deletion_multiple}
  Let $G=(V,E)$ be a graph and $p \in [0,1]$.
  Then, with probability $1-n^{-10}$ over $F\sim_p E$, we have
  \[
    \bm{x}^\top E_F \bm{x} \leq %
    \max\left\{6p\lambda_n(G),24\log n \right\}.
  \]
  for any unit vector $\bm{x} \in \mathbb{R}^n$.
\end{lemma}
\begin{proof}
  For an edge $e\in E$, let $X_e$ be the indicator random variable of the event that $e$ is included in $F$. Note that \[E_F=\sum_{e\in F}E_e=\sum_{e\in E}X_e \cdot E_e.\]

  By the fact that $\Pr[X_e=1]=p$, we have $\E[E_F]=\sum_{e\in E}pE_e=pL_G$.
  Let $\mu_{\max}:=\lambda_{\max}(\E[E_F])= p\lambda_{\max}(L_G)=p\lambda_n$.

  Note that the variables $X_e\;(e\in E)$ are independent and thus $E_F$ is a sum of independent random variables $X_e \cdot E_e$.
  Further note that $0\leq \lambda_{\min}(X_e \cdot E_e)\leq \lambda_{\max}(X_e \cdot E_e) \leq 1$.
  Now by the matrix Chernoff bound (Theorem~\ref{thm:matrix_chernoff}),
  for any $\varepsilon>0$, we have that
  \begin{eqnarray*}
  	\Pr[\lambda_{\max}(E_F)\geq (1+\varepsilon)\mu_{\max}]\leq n{\left(\frac{e^\varepsilon}{{(1+\varepsilon)}^{1+\varepsilon}}\right)}^{\mu_{\max}}
  \end{eqnarray*}

If $\mu_{\max}\leq 4\log n$, then by setting $\varepsilon= \frac{24\log n}{\mu_{\max}}-1\geq 2e-1$,
  \begin{eqnarray*}
  	\Pr[\lambda_{\max}(E_F)\geq (1+\varepsilon)\mu_{\max}]\leq n2^{-(1+\varepsilon)\mu_{\max}}	\leq \frac{1}{n^{10}}
  \end{eqnarray*}
  If $\mu_{\max}>4\log n$, then by setting $\varepsilon=5$, we have that
  \[\Pr[\lambda_{\max}(E_F)\geq (1+\varepsilon)\mu_{\max}]\leq n2^{-(1+\varepsilon)\mu_{\max}}	\leq \frac{1}{n^{10}} \]

  Thus with probability at least $1-n^{-10}$, %
  for any unit vector $\bm{x} \in \mathbb{R}^n$,
  \[
    \bm{x}^\top E_F \bm{x} \leq %
    \max\{6 \mu_{\max}, 24\log n\} =\max\left\{6p\lambda_n,24\log n \right\}.
    \qedhere
  \]
\end{proof}

\begin{proof}[Proof of Lemma~\ref{lem:eigenvalue-unnormalized}]
By Lemma~\ref{lemma:random_deletion_multiple}, Lemma~\ref{lem:eigenvalue_multiple} and the fact that $\lambda_n(G)\leq 2\Delta$, it holds that with probability at least $1-n^{-10}$, $\lambda_3(G-F)\geq \lambda_3(G)-\max\{12p \Delta, 24\log n \}$. Then, the
inequality in the statement of the lemma directly follows from Assumption~\ref{assume:unnormalized}(i).
\end{proof}

\subsection{Average Sensitivity of \texorpdfstring{$\USC_2$}{USC}}\label{subsec:spectral-clustering-unnormalized}
In this section, we prove Theorem~\ref{thm:spectraclustering}.

For a graph $G = (V,E)$, we say %
a set
$S^* \subseteq V$ is \emph{an optimum solution of $G$ with respect to cut ratio} if $\alpha_G(S^*) = \alpha(G)$ and $|S^*|\leq |V|/2$.
We first show the following.

\begin{lemma}\label{lemma:approx-unnormalized}
Suppose that Assumption~\ref{assume:unnormalized} holds. Let $F\sim_p E$. Let $S^*$ and $S_F^*$ be optimum solutions of $G$ and $G-F$ with respect to cut ratio, respectively. Then the following holds with probability at least $1-p_{\fail}$:
\begin{itemize}
\item if $G$ is not connected, then $S_F^*=S^*$ and $(S_F^*,\overline{S_F^*})$ is the unique cut with cut ratio $0$;
\item otherwise, then  $\lambda_2(G-F) > 0$, and
\[
\alpha_{G-F}(S_F^*)\leq \alpha_{G-F}(S^*) \leq \mathrm{APP} \cdot \alpha_{G-F}(S_F^*),
\]
where $\mathrm{APP} = O\left(\frac{\lambda_2(G)}{\lambda_2(G-F)}\sqrt{\frac{\Delta}{\lambda_3(G)}}\right)$.
\end{itemize}

\end{lemma}
\begin{proof}
We first consider the former case.  Since $\lambda_3(G)\geq \Omega(\log n)$, $S^*$ and $\overline{S^*}$ induce two connected components of $G$. By Assumption~\ref{assume:unnormalized}(iii), with probability $1-p_\fail$, $(G-F)[S^*]$ and $(G-F)[\overline{S^*}]$ are still connected. Thus $S_F^*=S^*$, which corresponds to the unique cut with cut ratio $0$.

Now we consider the latter case. By Assumption~\ref{assume:unnormalized}(iii), with probability $1-{p_{\fail}}$, the resulting graph $G-F$ is connected, i.e., $\lambda_2(G-F)>0$.

By definition of cut ratio and Lemma~\ref{lem:improvedCheeger}, it holds that $$\alpha_{G-F}(S_F^*)\leq \alpha_{G-F}(S^*)\leq \alpha_G(S^*)=O\left(\lambda_2(G)\sqrt{\frac{\Delta}{\lambda_3(G)}}\right).$$

Furthermore by Lemma~\ref{lem:Cheeger}, $\lambda_2(G-F)/2\leq\alpha(G-F)= \alpha_{G-F}(S_F^*)$ and thus we have
\[
  \alpha_{G-F}(S^*) = O\left(\frac{\lambda_2(G)}{\lambda_2(G-F)}\sqrt{\frac{\Delta}{\lambda_3(G)}} \cdot \alpha_{G-F}(S_F^*)\right).\qedhere
\]
\end{proof}

\begin{proof}[Proof of Theorem~\ref{thm:spectraclustering}]
Let $F\sim_p E$. Let $S^*$ and  $S_F^*$ be optimum solutions of $G$ and $G-F$ with respect to cut ratio, respectively. Let $S$ (resp. $S_F$) be the output of $\USC_2(G)$ (resp. $\USC_2(G-F)$). %
We further let $\mathcal{S}^*=\{S^*, \overline{S^*}\}$ be the bipartitioning corresponding to $S^*$. We define $\mathcal{S}^*_F, \mathcal{S}, \mathcal{S}_F$ similarly.

Let $\mathcal{E}$ denote the event that all the statements of Lemma~\ref{lem:eigenvalue-unnormalized} %
 and~\ref{lemma:approx-unnormalized} hold.
Then $\Pr[\mathcal{E}]\geq 1-n^{-10}-p_{\fail}\geq 1-2p_{\fail}$. We first assume that $\mathcal{E}$ holds.

In the case that $G$ is not connected, by Lemma~\ref{lem:Cheeger}, %
the partition $\mathcal{S}$ (resp. $\mathcal{S}_F$) is equivalent to $\mathcal{S}^*$ (resp. $\mathcal{S}_F^*$). Thus, by Lemma~\ref{lemma:approx-unnormalized}, $${d_{\mathrm{size}}(\USC_2(G),\USC_2(G-F))}=0.$$

Now, we assume that $G$ is connected. Let $\APP$ be the approximation ratio as specified in Lemma~\ref{lemma:approx-unnormalized}. Let
  \begin{align*}
    \varepsilon_1
    & :=\Theta\left(\frac{\Delta^{1/2}}{{\lambda_3(G)}^{1/2}} \cdot  \frac{\lambda_2(G) \Delta^{1/2}}{{\lambda_3(G)}^{3/2}}\right)
    = \Theta\left( \frac{\lambda_2(G)\Delta}{{\lambda_3(G)}^2}\right), \\
    \varepsilon_2
    & :=\Theta\left(\frac{\Delta^{1/2}}{{\lambda_3(G-F)}^{1/2}} \cdot \frac{\lambda_2(G-F) \Delta^{1/2}}{{\lambda_3(G-F)}^{3/2}}\right)
    = \Theta\left(\frac{ \lambda_2(G-F)\Delta}{{\lambda_3(G-F)}^2}\right), \\
    \varepsilon_3 &
    :=\Theta\left(\APP\cdot \frac{\lambda_2(G-F) \Delta^{1/2}}{{\lambda_3(G-F)}^{3/2}}\right)
    = \Theta\left( \frac{\lambda_2(G) \Delta}{\sqrt{\lambda_3(G){\lambda_3(G-F)}^{3}}}\right), \\
    \varepsilon_\star
    & =\max_{1\leq i\leq 3}\{\varepsilon_i\}
    =\Theta\left( \max\left\{\frac{\lambda_2(G)\Delta}{\sqrt{\lambda_3(G){\lambda_3(G-F)}^{3}}}, \frac{\lambda_2(G-F)\Delta}{{\lambda_3(G-F)}^2}\right\}\right).
  \end{align*}

  By Lemmas~\ref{lem:Cheeger} and~\ref{lem:improvedCheeger}, $S$ is an $O(\sqrt{\Delta/\lambda_3(G)})$-approximation of $S^*$.
  Thus, we have $d_{\mathrm{size}}\left(\mathcal{S},\mathcal{S}^*\right) \leq \varepsilon_1 n$ by Lemma~\ref{lemma:stable_kllot_Delta}.
  Similarly, $S_F$ is an $O(\sqrt{\Delta/\lambda_3(G-F)})$-approximation of $S_F^*$, and we have $d_{\mathrm{size}}(\mathcal{S}_F,\mathcal{S}_F^*) \leq \varepsilon_2 n$.

  By Lemma~\ref{lemma:approx-unnormalized}, $S^*$ is an $\APP$-approximation of $S_F^*$, and hence we have $d_{\mathrm{size}}(\mathcal{S}^*,\mathcal{S}_F^*) < \varepsilon_3 n$.
Since $\lambda_2(G-F)\leq \lambda_2(G)$ (by the monotone property of $\lambda_2$; see e.g., \cite{fiedler1973algebraic}),%
we have
  \begin{align*}
 &d_{\mathrm{size}}(\mathcal{S},\mathcal{S}_F)   \leq d_{\mathrm{size}}(\mathcal{S},\mathcal{S}^*)+d_{\mathrm{size}}(\mathcal{S}^*,\mathcal{S}_F^*)+d_{\mathrm{size}}(\mathcal{S}_F^*,\mathcal{S}_F) \\
 & \leq 3\varepsilon_\star n\leq \max\left\{\frac{\lambda_2(G)}{\sqrt{\lambda_3(G){\lambda_3(G-F)}^3}}, \frac{\lambda_2(G-F)}{{\lambda_3(G-F)}^2}\right\}\cdot O(\Delta n)\\
 &\leq O\left(\frac{\lambda_2(G)}{{\lambda_3(G)}^2}\cdot \Delta n\right)
  \end{align*}

  Then, we have
  \begin{align*}
  & \E_{F\sim_p E}[d_{\mathrm{size}}(\USC_2(G),\USC_2(G-F))] = \E_{F \sim_p E}[d_{\mathrm{size}}(\mathcal{S},\mathcal{S}_F)] \\
  &\leq O\left(\frac{\lambda_2(G)}{{\lambda_3(G)}^2}\cdot \Delta n \right) + n\cdot 2p_\fail = O\left(\frac{\lambda_2(G)}{{\lambda_3(G)}^2}\cdot \Delta n +1 \right),%
  \end{align*}
where in the inequality we used the fact that if $\mathcal{E}$ does not hold, then $d_{\mathrm{size}}(S,S_F)\leq n$.
\end{proof}

\section{Spectral Clustering with Normalized Laplacian}\label{sec:normalized}
In this section, we analyze the $p$-average sensitivity of $\SC_2$ (with respect to volume) for $p \in [0,1]$.
For a graph $G$, let $\nu_i(G)$ denote the $i$-th smallest eigenvalue of the normalized Laplacian associated with a graph $G$, that is, $\nu_i(G) = \lambda_i(\mathcal{L}_G)$.
The goal of this section is to show the following under Assumption~\ref{assume:normalized}, which we will explain in Section~\ref{subsec:assumption-normalized}.
\begin{theorem}\label{thm:normalspectraclustering}
	Let $G=(V,E)$ be a graph and $p\in [0,1]$.
	If Assumption~\ref{assume:normalized} holds, then the $p$-average sensitivity of $\SC_{2}$ with respect to volume is
	\[
		O\left(\frac{\nu_2(G)}{{\nu_3(G)}^2} \cdot \vol(G) + 1\right).%
	\]
\end{theorem}

We discuss Assumption~\ref{assume:normalized} and its plausibility in Section~\ref{subsec:assumption-normalized}, and then prove Theorem~\ref{thm:normalspectraclustering} in Section~\ref{subsec:spectral-clustering-normalized}.

\subsection{Assumptions}\label{subsec:assumption-normalized}
Let $G=(V,E)$ be a graph with minimum degree $\tau$ and maximum degree $\Delta$.
Recall that $C(p)$ is the reliability of $G$ given that each edge fails with probability $p$. Let $p_{\fail}':=O(\max\{\nu_2(G)/{\nu_3(G)}^2,{(2m)}^{-1}\})$.
\begin{assumption}\label{assume:normalized}
We assume the following properties hold.
	\begin{enumerate}[label=(\roman*)]
	\item[] \inlineitem $\nu_3(G)\geq \Omega(\tau^{-1} \log n)$;
	\inlineitem $p_{\fail}'\leq 1$;
	\item[] \inlineitem $p\leq O(\Delta^{-1}\log n)$ and $C(p)\geq 1-{p_{\fail}'}$.
	\end{enumerate}
\end{assumption}
\paragraph{Plausibility of Assumptions.} The Assumption~\ref{assume:normalized}(i), (ii) and (iii) can be justified similarly as in Section~\ref{subsec:assumption-unnormalized}. Note that (i) implicitly require that the minimum degree $\tau =\Omega(\log n)$, as $\nu_{3}(G)\leq 2$. %

\subsection{Average Sensitivity of \texorpdfstring{$\SC_{2}$}{NSC2}}\label{subsec:spectral-clustering-normalized}

The following gives a bound on the average sensitivity of eigenvalues of normalized Laplacian. 
We defer the proof to Appendix \ref{sec:proof-normalized_random_deletion_multiple}.
\begin{theorem}\label{thm:normalized_random_deletion_multiple}
	Let $G=(V,E)$ be a graph and $F \sim_p E$.
	If Assumption~\ref{assume:normalized}(i) and~(iii) hold, then we have $\nu_3(G-F)\geq \nu_3(G)/2$ with probability at least $1-n^{-7}$.
\end{theorem}
 Now we give the sketch of the proof of Theorem~\ref{thm:normalspectraclustering}.
\begin{proof}[Proof Sketch of Theorem~\ref{thm:normalspectraclustering}]
The proof is analogous to that in Section~\ref{subsec:spectral-clustering-unnormalized}. Here we mainly sketch the differences. %

We will consider the optimum solutions $S^*$ and  $S_F^*$ of $G$ and $G-F$ with respect to \emph{conductance}, respectively. Let $S$ (resp. $S_F$) be the output of $\SC_2(G)$ (resp. $\SC_2(G-F)$). We define $\varepsilon_1=\Theta\left( \frac{\nu_2(G)}{{\nu_3(G)}^2}\right),\varepsilon_2=\Theta\left(\frac{\nu_2(G-F)}{{\nu_3(G-F)}^2}\right), \varepsilon_3=\Theta\left( \frac{\nu_2(G)}{\sqrt{\nu_3(G){\nu_3(G-F)}^3}}\right)$, similarly as in the proof of Theorem~\ref{thm:spectraclustering}.  %

By Lemmas~\ref{lem:Cheeger},~\ref{lem:improvedCheeger}, and~\ref{lemma:stable_kllot_Delta}, for bipartitions $\mathcal{S}=\{S,\overline{S}\}$ and $\mathcal{S}^*=(S^*,\overline{S^*})$, it holds that  $d_{\mathrm{vol}}(\mathcal{S},\mathcal{S}^*) \leq \varepsilon_1 \vol(G)$.
For bipartitions $\mathcal{S}_F=\{S_F,\overline{S_F}\}$ and $\mathcal{S}_F^*=\{S^*_F,\overline{S^*_F}\}$, it holds that $d_{\mathrm{vol}}(\mathcal{S}_F,\mathcal{S}_F^*) \leq \varepsilon_2 \vol(G-F)\leq \varepsilon_2\vol(G)$.

Analogously to the proof of Lemma~\ref{lemma:approx-unnormalized}, we can show that $S^*$ is a good approximation of $S_F^*$ in $G-F$, and that $d_{\mathrm{vol}}(\mathcal{S}^*,\mathcal{S}_F^*) \leq \varepsilon_3 \vol(G)$.
By bounding the expectation as before, we can obtain the $p$-average sensitivity of $\SC_2$, as stated in the theorem.
\end{proof}

\section{Spectral Clustering with \texorpdfstring{$k$}{k} Clusters}\label{sec:kcluster}
In this section, we consider the $p$-average sensitivity of $\SC_{k\text{-means}}$.
For a graph $G$, let $\nu_i(G)$ denote the $i$-th smallest eigenvalue the normalized Laplacian $\mathcal{L}_G$.
We now prove the following.
\begin{theorem}\label{thm:kspectra}
	Let $G=(V,E)$ be a graph and $p\in [0,1]$.
	If Assumption~\ref{assume:k_cluster_normalized} holds, then the $p$-average sensitivity of $\SC_{k\text{-means}}$ with respect to volume is
	$$O\left(\frac{\alpha k^5\sqrt{\nu_k(G)}}{\nu_{k+1}(G)}\cdot \vol(G)+1\right),$$
	where $\alpha$ is the approximation ratio of $\appkmeans$.
\end{theorem}

\subsection{Assumptions}
Let $G=(V,E)$ be a graph with minimum degree $\tau$ and maximum degree $\Delta$.
Let $p_{\fail}'':=O(\max\{k^5\sqrt{\nu_{k}(G)}/{\nu_{k+1}(G)},{(2m)}^{-1}\})$.
\begin{assumption}\label{assume:k_cluster_normalized}
	We assume the following properties hold.
	\begin{enumerate}[label=(\roman*)]
		\item[] \inlineitem $\nu_{k+1}(G)\geq \Omega(\tau^{-1}\log n)$;
		\inlineitem $p_{\fail}''\leq 1$;
		\item[]  \inlineitem $p\leq O(\Delta^{-1}\log n)$ and $C(p)\geq 1-{p_{\fail}''}$;
		\item[] \inlineitem$\nu_{k+1}(G)/\rho_G(k)=\Omega(k^3)$.

	\end{enumerate}
\end{assumption}
The plausibility of the above assumptions can be justified almost the same as in Section~\ref{subsec:assumption-unnormalized} and~\ref{subsec:assumption-normalized}, except that we have one additional condition (iv), which further assumes that the input graph has a significant cluster structure, i.e., it has a $k$-way partition for which every cluster has low conductance.

\subsection{Proof of Theorem~\ref{thm:kspectra}}
Similar to the proof of Theorem~\ref{thm:normalized_random_deletion_multiple}, we have the following lemma regarding the perturbation of $\nu_{k+1}(G)$.
The only difference is that we use our new assumption on $\nu_{k+1}(G)$.
\begin{lemma}\label{lem:k_cluster_normalized_random_deletion_multiple}
	Let $G=(V,E)$ be a graph and $F \sim_p E$.
	If Assumption~\ref{assume:k_cluster_normalized}(i) and (iii) hold, then we have $\nu_{k+1}(G-F)\geq \frac{\nu_{k+1}(G)}{2}$ with probability at least $1-n^{-7}$.
\end{lemma}

We will make use the following lemma whose proof is provided in Appendix \ref{sec:proof-new_stable_kllot_Delta}.

\begin{lemma}\label{lemma:new_stable_kllot_Delta}
	Let $G=(V,E)$ be a graph.
	For any $c \geq 1$, $G$ is
	\begin{align*}
		\left(c,\Theta\left(\frac{ck^3 \rho_G(k)}{{\nu_{k+1}(G)}}\right)\right)\text{-stable}
	\end{align*}
	 with respect to conductance.
\end{lemma}

For a graph $G$, we say a $k$-partition  $\mathcal{S}^*=\{S_1^*,\ldots,S_k^*\}$ an \emph{optimum solution of $G$ with respect to $k$-way expansion}, if $\rho_G(\mathcal{S}^*)=\rho_G(k)$. Now we give the sketch of the proof of Theorem~\ref{thm:kspectra}.

\begin{figure*}[t!]
  \subfigure[SBM$2$]{\includegraphics[width=.33\hsize]{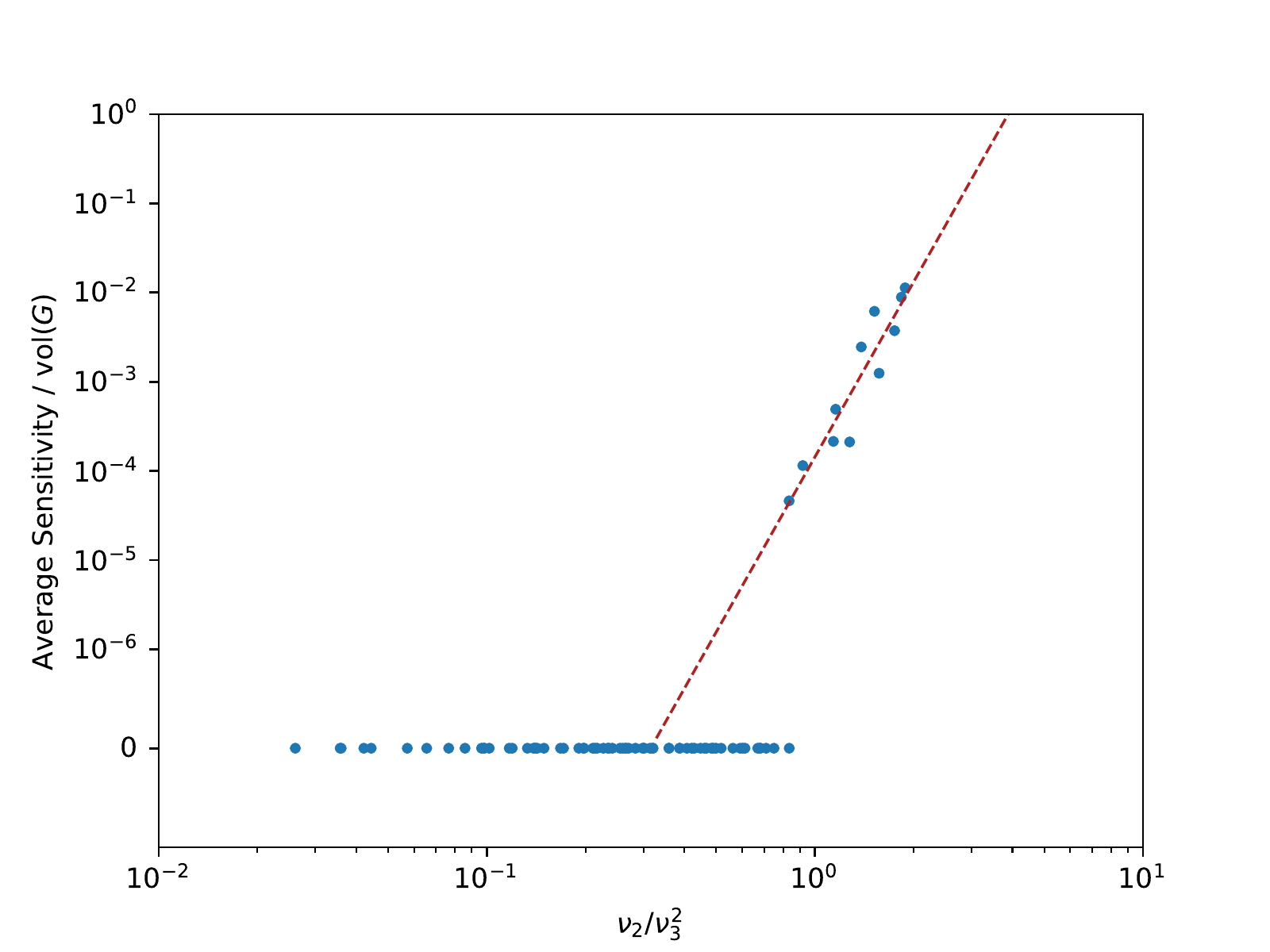}}
  \subfigure[LFR]{\includegraphics[width=.33\hsize]{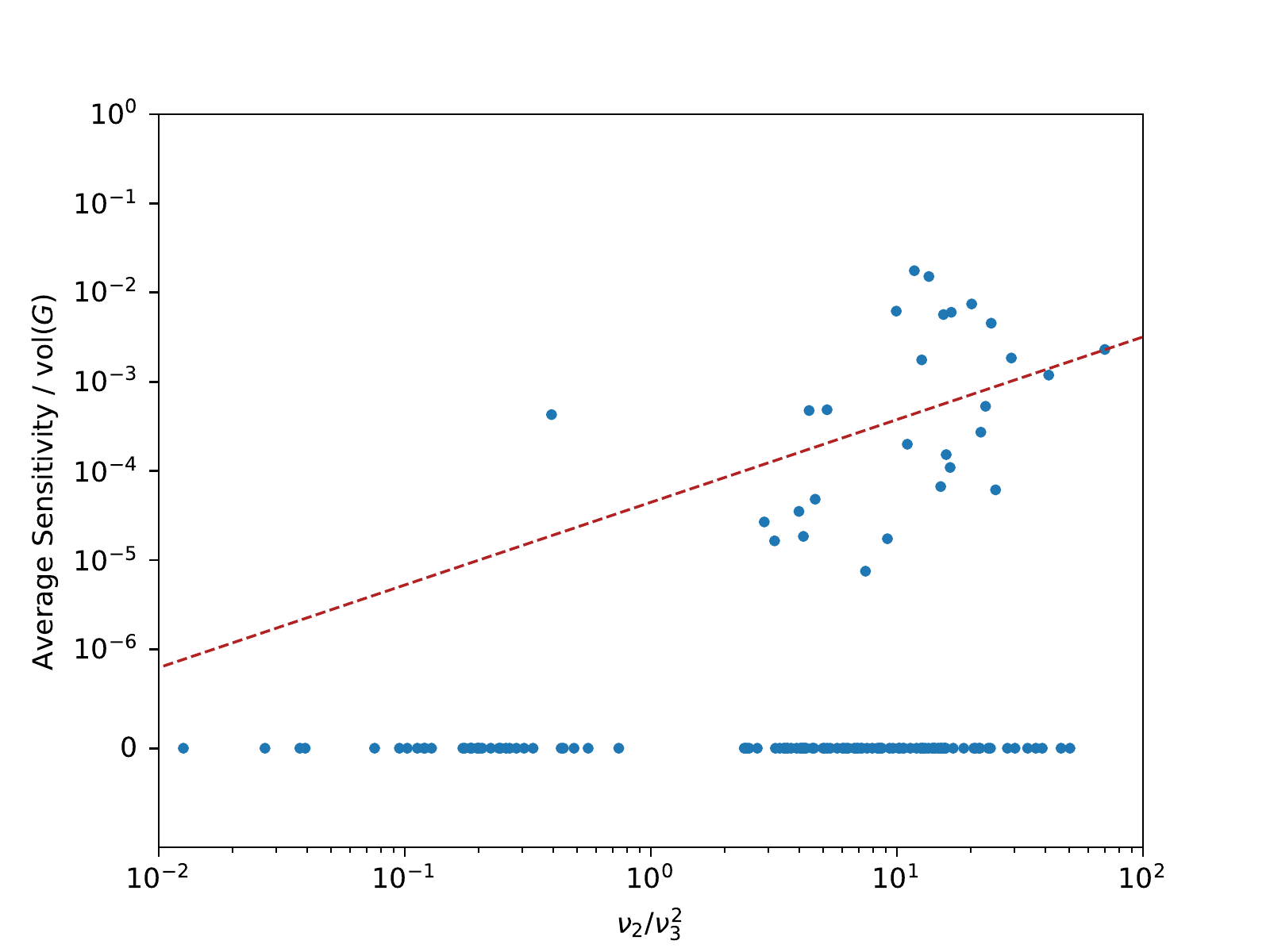}}
  \subfigure[Twitter]{\includegraphics[width=.33\hsize]{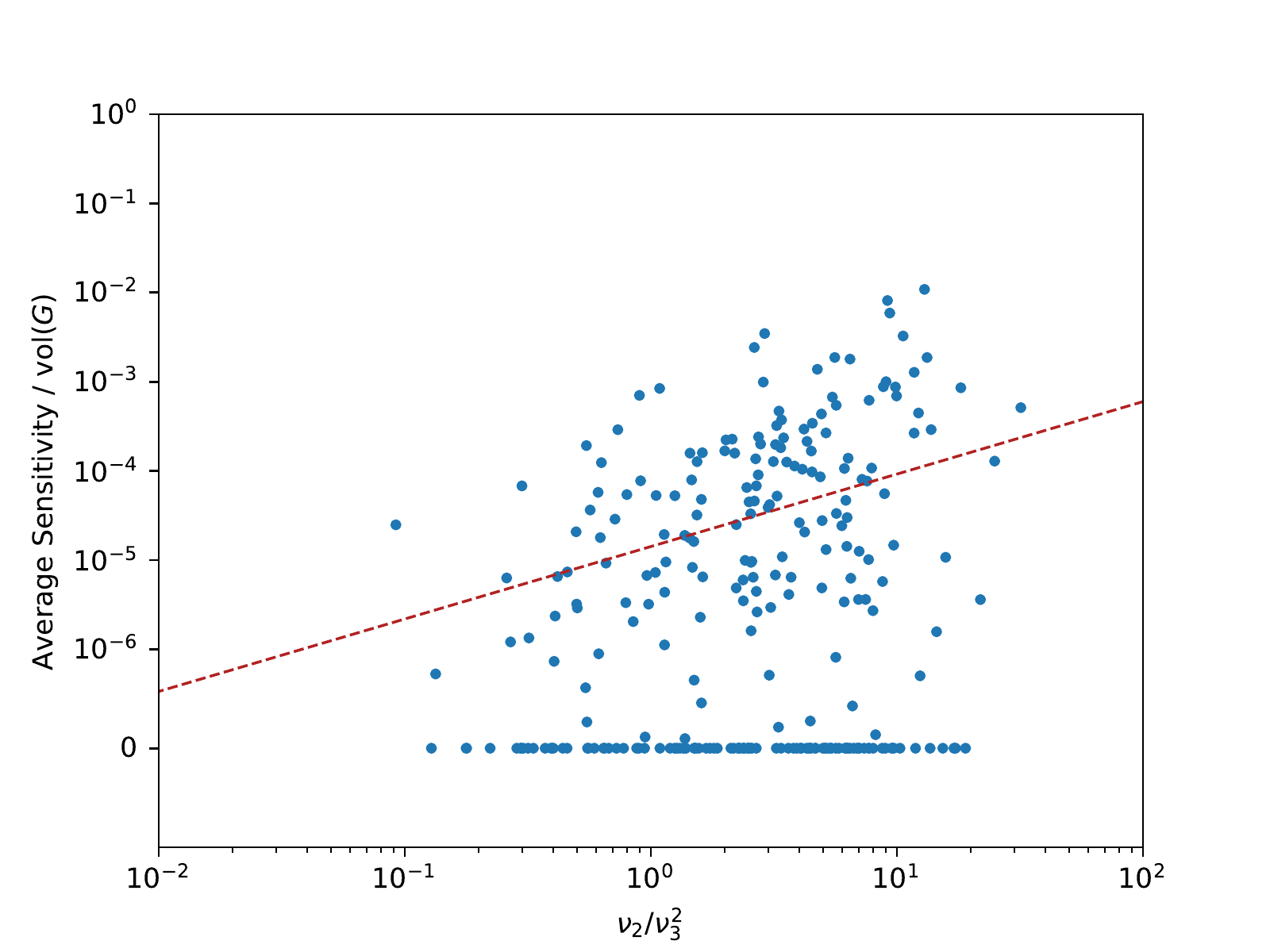}}
  \caption{Average sensitivity of $\SC_{2\text{-means}}$ and $\nu_{2}/\nu_{3}^2$. Each point represents a graph.}\label{fig:lambda-as}
\end{figure*}

\begin{proof}[Proof Sketch of Theorem~\ref{thm:kspectra}]
We will consider the optimum solutions $\mathcal{S}^*=\{S_1^*,\ldots,S_k^*\}$ and $\mathcal{S}_F^*=\{S_{1,F}^*,\ldots,S_{k,F}^*\}$ of $G$ and $G-F$ with respect to $k$-way expansion, respectively.
Let the $k$-partitions $\mathcal{S}=\{S_1,\ldots,S_k\}$ (resp. $\mathcal{S}_F=\{S_{1,F},\ldots,S_{k,F}\}$) be the output of $\SC_{k\text{-means}}(G)$ (resp. $\SC_{k\text{-means}}(G-F)$).

Let $\varepsilon_1:=\Theta\left(\alpha k^3 \cdot \frac{\rho_G(k)}{\nu_{k+1}(G)}\right)$. By the assumption that $\frac{\nu_{k+1}(G)}{\rho_G(k)}=\Omega(k^3)$, and Theorem~\ref{thm:goodkclustering}, $d_{\mathrm{vol}}\left(\mathcal{S},\mathcal{S}^*\right) \leq \varepsilon_1 \vol(G)$.

Let $\varepsilon_2:=\Theta\left(\alpha k^3 \cdot \frac{\rho_{G-F}(k)}{\nu_{k+1}(G-F)}\right)$.
By Lemma~\ref{lem:k_cluster_normalized_random_deletion_multiple}, the assumption $\nu_{k+1}(G)/\rho_G(k)=\Omega(k^3)$, and the fact that $\rho_{G-F}(k)\leq \rho_G(k)$ (as $\rho_G(k)$ is a monotone property), we have $\nu_{k+1}(G-F)/\rho_{G-F}(k)=\Omega(k^3)$. By Theorem~\ref{thm:goodkclustering}, $d_{\mathrm{vol}}\left(\mathcal{S}_F,\mathcal{S}_F^*\right) \leq \varepsilon_2 \vol(G)$.

Similarly to the proof of Lemma~\ref{lemma:approx-unnormalized}, by Assumption (iii) and (iv), we can show that with probability $1-p_\fail$, if $G$ contains $k$ connected components, then $\mathcal{S}_F^*=\mathcal{S}^*$; and otherwise, $\nu_{k}(G-F)>0$ and
\[
	\rho_{G-F}(\mathcal{S}_F^*)\leq \rho_{G-F}(\mathcal{S}^*) \leq \APP\cdot \rho_{G-F}(\mathcal{S}_F^*),
\]
where $\APP=\frac{\rho_{G}(\mathcal{S}^*)}{\rho_{G-F}(\mathcal{S}_F^*)}=\frac{\rho_{G}(k)}{\rho_{G-F}(k)}$. %
That is, $\mathcal{S}^*$ is an $\APP$-approximation of $\mathcal{S}_F^*$ in $G-F$.

Thus, we can set $\varepsilon_3=\Theta\left(k^3\frac{\rho_{G}(k)}{\rho_{G-F}(k)} \cdot \frac{\rho_{G-F}(k)}{\nu_{k+1}(G-F)}\right)=\Theta\left(\frac{k^3\rho_{G}(k)}{\nu_{k+1}(G-F)}\right)$.  %
By Lemma~\ref{lemma:new_stable_kllot_Delta}, $d_{\mathrm{vol}}\left(\mathcal{S}^*,\mathcal{S}_F^*\right) \leq \varepsilon_3 \vol(G)$. Thus
\begin{align*}
&d_{\mathrm{vol}}\left(\SC_{k\text{-means}}(G),\SC_{k\text{-means}}(G-F)\right)\\
&=d_{\mathrm{vol}}\left(\mathcal{S},\mathcal{S}_F\right) \leq (3\max_{1\leq i\leq 3}\varepsilon_i)\cdot \vol(G)\\
&=O\left(\frac{\alpha k^3\rho_{G}(k)}{\nu_{k+1}(G)}\vol(G)+1\right)=O\left( \frac{\alpha k^5\sqrt{\nu_k(G)}}{\nu_{k+1}(G)}\cdot \vol(G)+1\right).
\end{align*}

Finally, by bounding the expectation as before, we can obtain the $p$-average sensitivity of $\SC_{k\text{-means}}$ as stated in the theorem.
\end{proof}

\section{Experiments}\label{sec:experiments}
\begin{table}[t!]
  \caption{Datasets: $\#$, $\overline{n}$, $\overline{m}$, $\overline{\nu_2}$, $\overline{\nu_3}$ are the number of graphs, the average number of vertices, the average number of edges, the average of the second smallest eigenvalue (of the normalized Laplacian), the average of the third smallest eigenvalue, respectively.}\label{tab:datasets}
  \begin{tabular}{lrrrrr}
  \toprule
  Name & $\#$ & $\overline{n}$ & $\overline{m}$ & $\overline{\nu_2}$ & $\overline{\nu_3}$ \\
  \midrule
  SBM2 & 80 & 100.00 & 1532.04 & 0.183 & 0.713 \\
  SBM3 & 80 & 100.00 & 1122.86 & 0.216 & 0.239 \\
  SBM4 & 80 & 100.00 & 918.49 & 0.233 & 0.259 \\
  LFR & 170 & 100.00 & 1173.62 & 0.045 & 0.165 \\
  Twitter & 273 & 131.12 & 1684.22 & 0.057 & 0.169 \\
  \bottomrule
  \end{tabular}
\end{table}

In this section, we show our experimental results to validate our theoretical results.
Here, we focus on spectral clustering with normalized Laplacian ($\SC_{k\text{-means}}$) because it is advocated for practical use, as we mentioned in Section~\ref{subsubsec:spectral-clustering}. We obtained similar results for spectral clustering with unnormalized Laplacian.

As it is computationally hard to calculate the exact value of average sensitivity, we took the average of 1000 trials, where each trial samples a set of edges $F \sim_p E$ and removes $F$ from the graph to compute the symmetric difference size.
Also in our plots, we divided the average sensitivity by $\mathrm{vol}(G)$ so that we can compare graphs with different sizes.
We set edge removal probability $p$ to be $10^{-3}$ in all the experiments.

\vspace{-1em}

\subsection{Datasets}
In our experiments, we study five datasets, SBM2, SBM3, SBM4, LFR, and Twitter, which are explained below.

For $k\in \{2,3,4\}$, the SBM$k$ dataset is a collection of graphs with $k$ clusters generated with the \emph{stochastic block model}.
Specifically, we generate graphs of $100$ vertices with $k$ equal-sized clusters by adding an edge for each vertex pair within a cluster with probability $p$ and adding an edge for each vertex pair between different clusters with probability $q$ for each choice of $p \in \{0.3,0.4,\ldots,0.9\}$ and $q \in \{0.01,0.02,\ldots,0.1\}$.

The LFR dataset is a collection of graphs generated with the Lancichinetti-Fortunato-Radicchi (LFR) benchmark~\cite{Lancichinetti2008}.
We run the implementation provided by the authors\footnote{\url{https://www.santofortunato.net/resources}} to generate graphs with $100$ vertices, average degree $d$, the maximum degree $50$, and the mixing parameter $\mu$ for each choice of $d \in \{4,5,\ldots,20\}$ and $\mu \in \{0.01,0.02,\ldots,0.1\}$.

The Twitter dataset is a collection of ego-networks in the Twitter network provided at SNAP\footnote{\url{http://snap.stanford.edu/index.html}}.
As the original dataset was a collection of directed graphs, we discarded directions of edges.

We provide basic information about these datasets in Table~\ref{tab:datasets}.

\subsection{Results}
\paragraph{Average sensitivity of $2$-way clustering}
Figure~\ref{fig:lambda-as} shows the relation between $p$-average sensitivity and $\nu_2/\nu_3^2$, where each point represents a graph in the corresponding dataset.
The red lines were computed by applying linear regression on graphs with positive average sensitivity.
For the SBM$2$ dataset, we can observe a clear phase transition phenomenon: The average sensitivity dramatically increases when $\nu_2/\nu_{3}^2$ approaches to one.
In all the datasets, we can observe that average sensitivity increases as $\nu_2/\nu_{3}^2$ increases.
These results empirically confirm the validity of Theorem~\ref{thm:normalspectraclustering}.

\begin{figure}[t!]
  \subfigure[SBM$3$]{\includegraphics[width=.49\hsize]{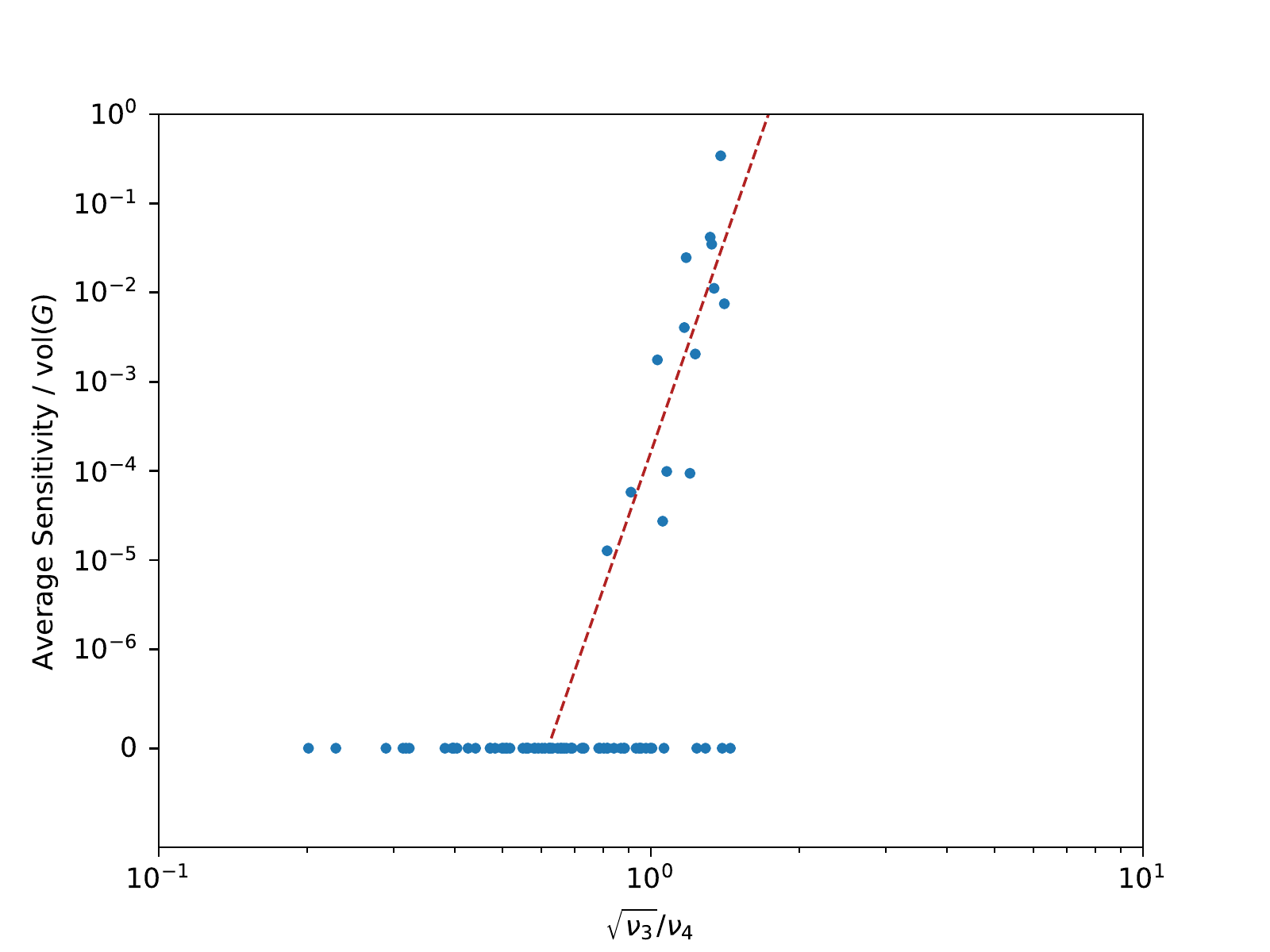}}
  \subfigure[SBM$4$]{\includegraphics[width=.49\hsize]{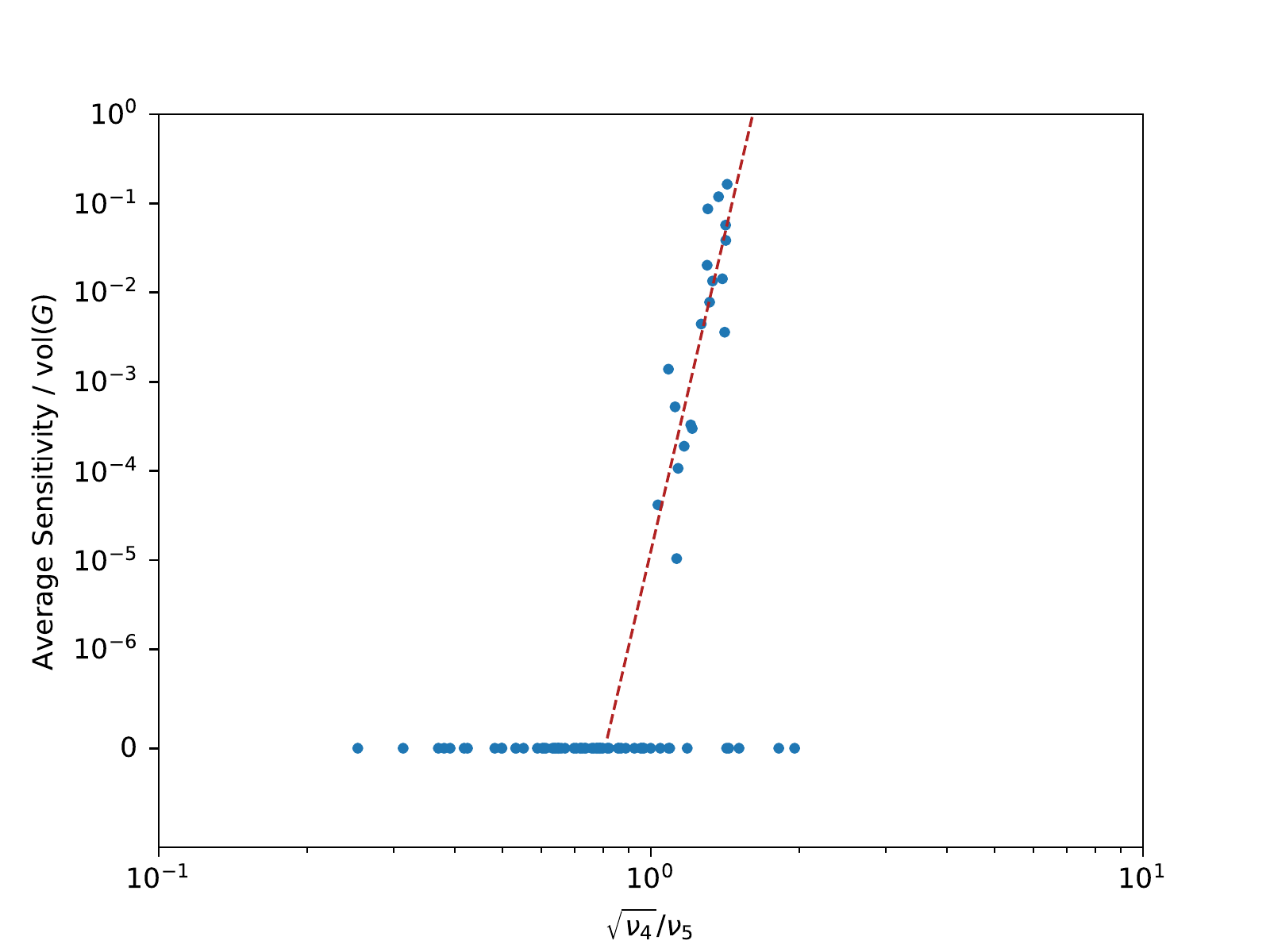}}
  \caption{Average sensitivity of $\SC_{k\text{-means}}$ and $\sqrt{\nu_{k}}/\nu_{k+1}$ on the SBM$k$ dataset. Each point represents a graph.}\label{fig:lambda-as-k}
\end{figure}

\paragraph{Average sensitivity of $k$-way clustering}
Figure~\ref{fig:lambda-as-k} shows the relation between the average sensitivity of $\SC_{k\mathrm{-means}}$ and $\sqrt{\nu_k}/\nu_{k+1}$ on the SBM$k$ datasets, which are collections of graphs with $k$ clusters.
As with the results for the SBM$2$ dataset in Figure~\ref{fig:lambda-as}, we can again observe a phase transition phenomenon.
These results suggest that the parameter $\sqrt{\nu_k}/\nu_{k+1}$ is critical for the average sensitivity of spectral clustering, as indicated in Theorem~\ref{thm:kspectra}.

\begin{figure}[t!]
  \begin{minipage}[t]{0.47\hsize}
  \includegraphics[width=\hsize]{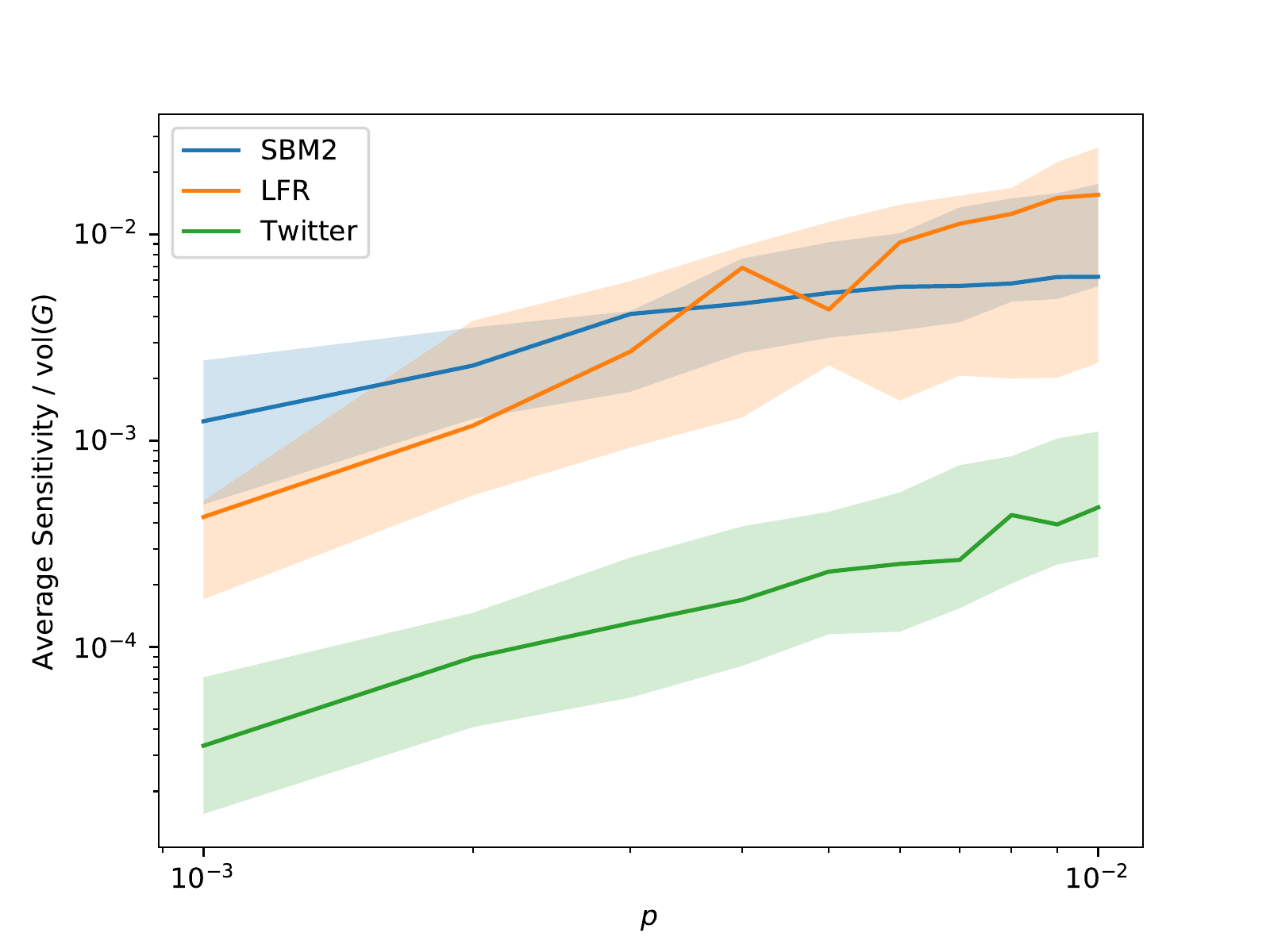}
  \caption{Average sensitivity of $\SC_{2\text{-means}}$ and edge removal probability $p$.}\label{fig:k-as}
  \end{minipage}
  \quad
  \begin{minipage}[t]{0.47\hsize}
  \includegraphics[width=\hsize]{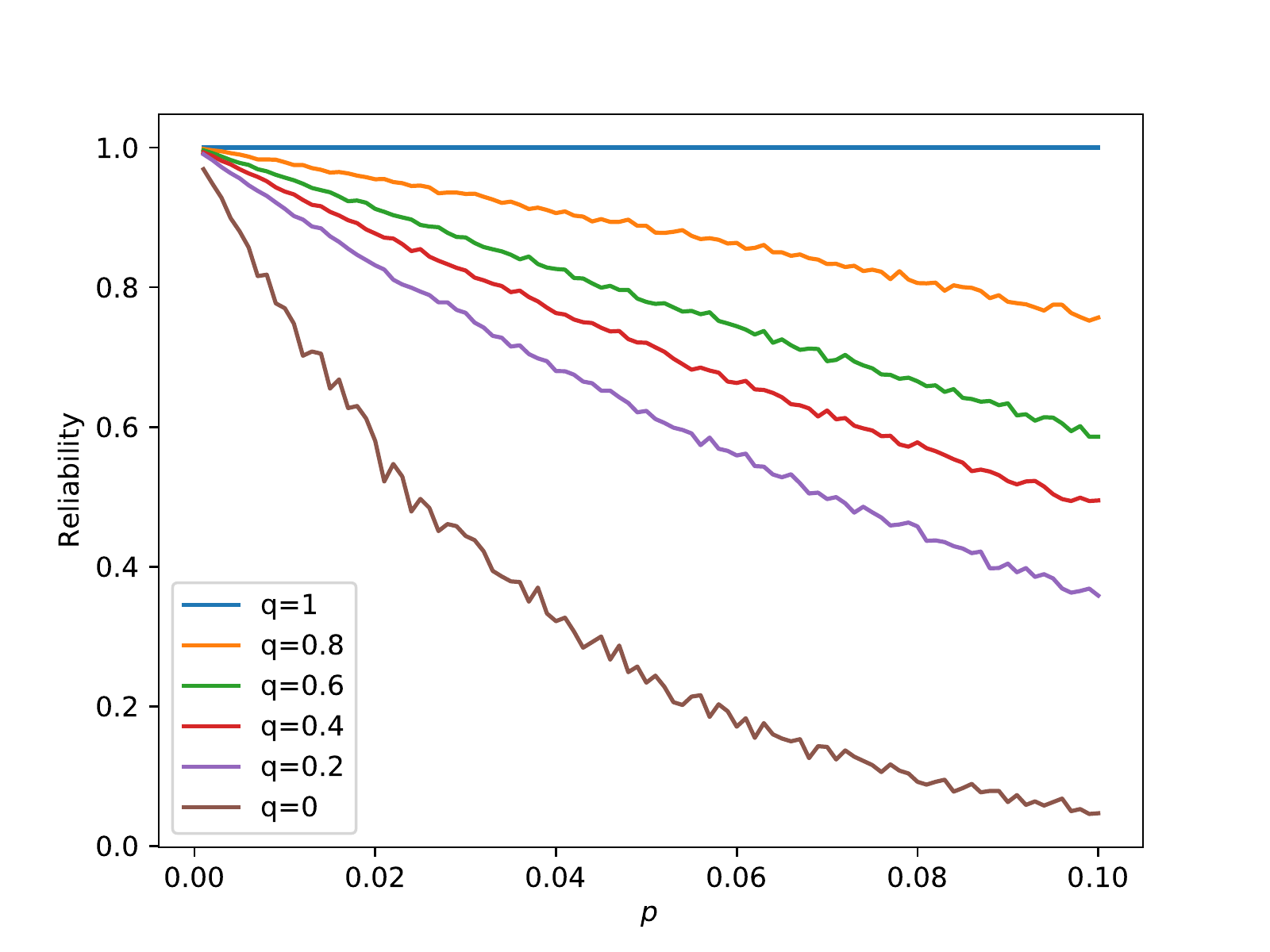}
  \caption{Quantile curves of reliability of graphs in the Twitter dataset.}\label{fig:reliability}
  \end{minipage}
\end{figure}

\paragraph{Average sensitivity and edge removal probability}
Figure~\ref{fig:k-as} shows the relation between average sensitivity grows and edge removal probability $p$, where a bold line shows the median of average sensitivities of graphs in the corresponding dataset, and the top and bottom of the filled region show the $0.6$-quantile and $0.4$-quantile, respectively.
We can observe that average sensitivity grows almost linear in $p$.
Such a linearity relation has been implicit in our theoretical analysis. For example, under our Assumption~\ref{assume:k_cluster_normalized}, the proof of Lemma~\ref{lem:k_cluster_normalized_random_deletion_multiple} (which is similar to the proof of Theorem~\ref{thm:normalized_random_deletion_multiple}) actually gives $\nu_{k+1}(G-F)\geq {\nu_{k+1}(G)} - 4 p$ for $p=O(\Delta^{-1}\log n)$ and $\nu_{k+1}(G)\geq \Omega(\tau^{-1}\log n)$. Furthermore, the proof of Theorem~\ref{thm:kspectra} implies that for small enough $p$, the average sensitivity of $\SC_{k\text{-means}}$ is $O\left(\frac{\alpha k^5\sqrt{\nu_k(G)}}{\nu_{k+1}(G-F)}\cdot \vol(G)+1\right)=O\left(\frac{\alpha k^5\sqrt{\nu_k}}{\nu_{k+1}}\cdot \vol(G)\left(1+ \frac{5p}{\nu_{k+1}}\right) +1\right)$, which is linear in $p$.

\paragraph{Reliability}
To confirm our assumptions (Assumptions~\ref{assume:unnormalized},~\ref{assume:normalized}, and~\ref{assume:k_cluster_normalized}) hold in practice, for each edge failure probability $p \in \{0.001,0.002,\ldots,0.1\}$, we computed the reliability of the 273 graphs in the Twitter dataset and calculated the $q$-quantile of the reliabilities for each $q \in \{0.01,0.02,\ldots,1\}$.
Figure~\ref{fig:reliability} shows the results.
As we can observe, most of the graphs in the Twitter dataset are connected after removing a 10\% of edges with high probability, which confirm the plausibility of the assumptions.

\section{Conclusions}\label{sec:conclusions}
To make the decision process more reliable and efficient, we initiate the study of the stability of spectral clustering by using the notion of average sensitivity.
We showed that $2$-way spectral clustering with both unnormalized and normalized Laplacians has average sensitivity proportional to $\lambda_2/\lambda_3^2$, and $k$-way spectral clustering with normalized Laplacian has average sensitivity proportional to $\sqrt{\lambda_k}/\lambda_{k+1}$, where $\lambda_i$ is the $i$-th smallest eigenvalue of the corresponding Laplacian.
We empirically confirmed these theoretical bounds using synthetic and real networks.
These results imply that we can reliably use spectral clustering because it is stable against random edge perturbations if there is a significant cluster structure in a graph.

\section*{Acknowledgments}
Y.~Y. is supported by was supported by JST, PRESTO Grant Number JPMJPR192B, Japan.

\bibliographystyle{ACM-Reference-Format}
\bibliography{sensispectra}


\begin{thebibliography}{26}


\ifx \showCODEN    \undefined \def \showCODEN     #1{\unskip}     \fi
\ifx \showDOI      \undefined \def \showDOI       #1{#1}\fi
\ifx \showISBNx    \undefined \def \showISBNx     #1{\unskip}     \fi
\ifx \showISBNxiii \undefined \def \showISBNxiii  #1{\unskip}     \fi
\ifx \showISSN     \undefined \def \showISSN      #1{\unskip}     \fi
\ifx \showLCCN     \undefined \def \showLCCN      #1{\unskip}     \fi
\ifx \shownote     \undefined \def \shownote      #1{#1}          \fi
\ifx \showarticletitle \undefined \def \showarticletitle #1{#1}   \fi
\ifx \showURL      \undefined \def \showURL       {\relax}        \fi
\providecommand\bibfield[2]{#2}
\providecommand\bibinfo[2]{#2}
\providecommand\natexlab[1]{#1}
\providecommand\showeprint[2][]{arXiv:#2}

\bibitem[\protect\citeauthoryear{Alon}{Alon}{1986}]%
        {Alon:1986gz}
\bibfield{author}{\bibinfo{person}{Noga Alon}.}
  \bibinfo{year}{1986}\natexlab{}.
\newblock \showarticletitle{Eigenvalues and expanders}.
\newblock \bibinfo{journal}{\emph{Combinatorica}} \bibinfo{volume}{6},
  \bibinfo{number}{2} (\bibinfo{year}{1986}), \bibinfo{pages}{83--96}.
\newblock


\bibitem[\protect\citeauthoryear{Alon and Milman}{Alon and Milman}{1985}]%
        {Alon:1985jg}
\bibfield{author}{\bibinfo{person}{Noga Alon} {and} \bibinfo{person}{V~D
  Milman}.} \bibinfo{year}{1985}\natexlab{}.
\newblock \showarticletitle{$\lambda_1$, Isoperimetric inequalities for graphs,
  and superconcentrators}.
\newblock \bibinfo{journal}{\emph{Journal of Combinatorial Theory, Series B}}
  \bibinfo{volume}{38}, \bibinfo{number}{1} (\bibinfo{year}{1985}),
  \bibinfo{pages}{73--88}.
\newblock


\bibitem[\protect\citeauthoryear{Belkin and Niyogi}{Belkin and Niyogi}{2001}]%
        {Belkin2001}
\bibfield{author}{\bibinfo{person}{Mikhail Belkin} {and}
  \bibinfo{person}{Partha Niyogi}.} \bibinfo{year}{2001}\natexlab{}.
\newblock \showarticletitle{Laplacian Eigenmaps and Spectral Techniques for
  Embedding and Clustering}. In \bibinfo{booktitle}{\emph{NIPS}}.
  \bibinfo{pages}{585--591}.
\newblock


\bibitem[\protect\citeauthoryear{Bilu and Linial}{Bilu and Linial}{2012}]%
        {BL12:stable}
\bibfield{author}{\bibinfo{person}{Yonatan Bilu} {and} \bibinfo{person}{Nathan
  Linial}.} \bibinfo{year}{2012}\natexlab{}.
\newblock \showarticletitle{Are stable instances easy?}
\newblock \bibinfo{journal}{\emph{Combinatorics, Probability and Computing}}
  \bibinfo{volume}{21}, \bibinfo{number}{5} (\bibinfo{year}{2012}),
  \bibinfo{pages}{643--660}.
\newblock


\bibitem[\protect\citeauthoryear{Colbourn}{Colbourn}{1987}]%
        {colbourn1987combinatorics}
\bibfield{author}{\bibinfo{person}{Charles~J Colbourn}.}
  \bibinfo{year}{1987}\natexlab{}.
\newblock \bibinfo{booktitle}{\emph{The combinatorics of network reliability}}.
\newblock \bibinfo{publisher}{Oxford University Press, Inc.}
\newblock


\bibitem[\protect\citeauthoryear{Eldridge, Belkin, and Wang}{Eldridge
  et~al\mbox{.}}{2018}]%
        {eldridge2018unperturbed}
\bibfield{author}{\bibinfo{person}{Justin Eldridge}, \bibinfo{person}{Mikhail
  Belkin}, {and} \bibinfo{person}{Yusu Wang}.} \bibinfo{year}{2018}\natexlab{}.
\newblock \showarticletitle{Unperturbed: spectral analysis beyond Davis-Kahan}.
  In \bibinfo{booktitle}{\emph{Algorithmic Learning Theory}}.
  \bibinfo{pages}{321--358}.
\newblock


\bibitem[\protect\citeauthoryear{Fiedler}{Fiedler}{1973}]%
        {fiedler1973algebraic}
\bibfield{author}{\bibinfo{person}{Miroslav Fiedler}.}
  \bibinfo{year}{1973}\natexlab{}.
\newblock \showarticletitle{Algebraic connectivity of graphs}.
\newblock \bibinfo{journal}{\emph{Czechoslovak mathematical journal}}
  \bibinfo{volume}{23}, \bibinfo{number}{2} (\bibinfo{year}{1973}),
  \bibinfo{pages}{298--305}.
\newblock


\bibitem[\protect\citeauthoryear{Fortunato}{Fortunato}{2010}]%
        {Fortunato2010}
\bibfield{author}{\bibinfo{person}{Santo Fortunato}.}
  \bibinfo{year}{2010}\natexlab{}.
\newblock \showarticletitle{Community detection in graphs}.
\newblock \bibinfo{journal}{\emph{Physics Reports}} \bibinfo{volume}{486},
  \bibinfo{number}{3-5} (\bibinfo{year}{2010}), \bibinfo{pages}{75--174}.
\newblock


\bibitem[\protect\citeauthoryear{Gfeller, Chappelier, and De~Los~Rios}{Gfeller
  et~al\mbox{.}}{2005}]%
        {gfeller2005finding}
\bibfield{author}{\bibinfo{person}{David Gfeller},
  \bibinfo{person}{Jean-C{\'e}dric Chappelier}, {and} \bibinfo{person}{Paolo
  De~Los~Rios}.} \bibinfo{year}{2005}\natexlab{}.
\newblock \showarticletitle{Finding instabilities in the community structure of
  complex networks}.
\newblock \bibinfo{journal}{\emph{Physical Review E}} \bibinfo{volume}{72},
  \bibinfo{number}{5} (\bibinfo{year}{2005}), \bibinfo{pages}{056135}.
\newblock


\bibitem[\protect\citeauthoryear{Guo and Jerrum}{Guo and Jerrum}{2019}]%
        {GJ19:polynomial}
\bibfield{author}{\bibinfo{person}{Heng Guo} {and} \bibinfo{person}{Mark
  Jerrum}.} \bibinfo{year}{2019}\natexlab{}.
\newblock \showarticletitle{A polynomial-time approximation algorithm for
  all-terminal network reliability}.
\newblock \bibinfo{journal}{\emph{SIAM J. Comput.}} \bibinfo{volume}{48},
  \bibinfo{number}{3} (\bibinfo{year}{2019}), \bibinfo{pages}{964--978}.
\newblock


\bibitem[\protect\citeauthoryear{Hoory, Linial, and Wigderson}{Hoory
  et~al\mbox{.}}{2006}]%
        {hoory2006expander}
\bibfield{author}{\bibinfo{person}{Shlomo Hoory}, \bibinfo{person}{Nathan
  Linial}, {and} \bibinfo{person}{Avi Wigderson}.}
  \bibinfo{year}{2006}\natexlab{}.
\newblock \showarticletitle{Expander graphs and their applications}.
\newblock \bibinfo{journal}{\emph{Bull. Amer. Math. Soc.}}
  \bibinfo{volume}{43}, \bibinfo{number}{4} (\bibinfo{year}{2006}),
  \bibinfo{pages}{439--561}.
\newblock


\bibitem[\protect\citeauthoryear{Huang, Yan, Jordan, and Taft}{Huang
  et~al\mbox{.}}{2008}]%
        {Huang2008}
\bibfield{author}{\bibinfo{person}{Ling Huang}, \bibinfo{person}{Donghui Yan},
  \bibinfo{person}{Michael~I. Jordan}, {and} \bibinfo{person}{Nina Taft}.}
  \bibinfo{year}{2008}\natexlab{}.
\newblock \showarticletitle{Spectral Clustering with Perturbed Data}. In
  \bibinfo{booktitle}{\emph{NIPS}}. \bibinfo{pages}{705--712}.
\newblock


\bibitem[\protect\citeauthoryear{Karrer, Levina, and Newman}{Karrer
  et~al\mbox{.}}{2008}]%
        {karrer2008robustness}
\bibfield{author}{\bibinfo{person}{Brian Karrer}, \bibinfo{person}{Elizaveta
  Levina}, {and} \bibinfo{person}{Mark~EJ Newman}.}
  \bibinfo{year}{2008}\natexlab{}.
\newblock \showarticletitle{Robustness of community structure in networks}.
\newblock \bibinfo{journal}{\emph{Physical review E}} \bibinfo{volume}{77},
  \bibinfo{number}{4} (\bibinfo{year}{2008}), \bibinfo{pages}{046119}.
\newblock


\bibitem[\protect\citeauthoryear{Kato}{Kato}{2013}]%
        {kato2013perturbation}
\bibfield{author}{\bibinfo{person}{Tosio Kato}.}
  \bibinfo{year}{2013}\natexlab{}.
\newblock \bibinfo{booktitle}{\emph{Perturbation theory for linear operators}}.
  Vol.~\bibinfo{volume}{132}.
\newblock \bibinfo{publisher}{Springer Science \& Business Media}.
\newblock


\bibitem[\protect\citeauthoryear{Kwok, Lau, Lee, Oveis~Gharan, and
  Trevisan}{Kwok et~al\mbox{.}}{2013a}]%
        {KLLOT:improved}
\bibfield{author}{\bibinfo{person}{Tsz~Chiu Kwok}, \bibinfo{person}{Lap~Chi
  Lau}, \bibinfo{person}{Yin~Tat Lee}, \bibinfo{person}{Shayan Oveis~Gharan},
  {and} \bibinfo{person}{Luca Trevisan}.} \bibinfo{year}{2013}\natexlab{a}.
\newblock \showarticletitle{Improved Cheeger's inequality: Analysis of spectral
  partitioning algorithms through higher order spectral gap}. In
  \bibinfo{booktitle}{\emph{STOC}}. \bibinfo{pages}{11--20}.
\newblock


\bibitem[\protect\citeauthoryear{Kwok, Lau, Lee, Oveis~Gharan, and
  Trevisan}{Kwok et~al\mbox{.}}{2013b}]%
        {KLLOT:arxiv}
\bibfield{author}{\bibinfo{person}{Tsz~Chiu Kwok}, \bibinfo{person}{Lap~Chi
  Lau}, \bibinfo{person}{Yin~Tat Lee}, \bibinfo{person}{Shayan Oveis~Gharan},
  {and} \bibinfo{person}{Luca Trevisan}.} \bibinfo{year}{2013}\natexlab{b}.
\newblock \showarticletitle{Improved Cheeger's Inequality: Analysis of Spectral
  Partitioning Algorithms through Higher Order Spectral Gap}.
\newblock \bibinfo{journal}{\emph{arXiv preprint arXiv:1301.5584}}
  (\bibinfo{year}{2013}).
\newblock


\bibitem[\protect\citeauthoryear{Lancichinetti, Fortunato, and
  Radicchi}{Lancichinetti et~al\mbox{.}}{2008}]%
        {Lancichinetti2008}
\bibfield{author}{\bibinfo{person}{Andrea Lancichinetti},
  \bibinfo{person}{Santo Fortunato}, {and} \bibinfo{person}{Filippo Radicchi}.}
  \bibinfo{year}{2008}\natexlab{}.
\newblock \showarticletitle{Benchmark graphs for testing community detection
  algorithms}.
\newblock \bibinfo{journal}{\emph{Physical Review E}} \bibinfo{volume}{78},
  \bibinfo{number}{4} (\bibinfo{year}{2008}).
\newblock


\bibitem[\protect\citeauthoryear{Lee, {Oveis Gharan}, and Trevisan}{Lee
  et~al\mbox{.}}{2014}]%
        {lee2014multiway}
\bibfield{author}{\bibinfo{person}{James~R Lee}, \bibinfo{person}{Shayan {Oveis
  Gharan}}, {and} \bibinfo{person}{Luca Trevisan}.}
  \bibinfo{year}{2014}\natexlab{}.
\newblock \showarticletitle{Multiway spectral partitioning and higher-order
  cheeger inequalities}.
\newblock \bibinfo{journal}{\emph{J. ACM}} \bibinfo{volume}{61},
  \bibinfo{number}{6} (\bibinfo{year}{2014}), \bibinfo{pages}{37}.
\newblock


\bibitem[\protect\citeauthoryear{MacQueen}{MacQueen}{1967}]%
        {MacQueen1967}
\bibfield{author}{\bibinfo{person}{J. MacQueen}.}
  \bibinfo{year}{1967}\natexlab{}.
\newblock \showarticletitle{Some methods for classification and analysis of
  multivariate observations}. In \bibinfo{booktitle}{\emph{Proceedings of the
  5th Berkeley Symposium on Mathematical Statistics and Probability}}.
  \bibinfo{pages}{Vol. I: Statistics, pp. 281--297}.
\newblock


\bibitem[\protect\citeauthoryear{Peng, Sun, and Zanetti}{Peng
  et~al\mbox{.}}{2017}]%
        {peng2017partitioning}
\bibfield{author}{\bibinfo{person}{Richard Peng}, \bibinfo{person}{He Sun},
  {and} \bibinfo{person}{Luca Zanetti}.} \bibinfo{year}{2017}\natexlab{}.
\newblock \showarticletitle{Partitioning Well-Clustered Graphs: Spectral
  Clustering Works!}
\newblock \bibinfo{journal}{\emph{SIAM J. Comput.}} \bibinfo{volume}{46},
  \bibinfo{number}{2} (\bibinfo{year}{2017}), \bibinfo{pages}{710--743}.
\newblock


\bibitem[\protect\citeauthoryear{Shi and Malik}{Shi and Malik}{2000}]%
        {Shi2000}
\bibfield{author}{\bibinfo{person}{Jianbo Shi} {and} \bibinfo{person}{J.
  Malik}.} \bibinfo{year}{2000}\natexlab{}.
\newblock \showarticletitle{Normalized cuts and image segmentation}.
\newblock \bibinfo{journal}{\emph{{IEEE} Transactions on Pattern Analysis and
  Machine Intelligence}} \bibinfo{volume}{22}, \bibinfo{number}{8}
  (\bibinfo{year}{2000}), \bibinfo{pages}{888--905}.
\newblock


\bibitem[\protect\citeauthoryear{Stewart and Sun}{Stewart and Sun}{1990}]%
        {stewart1990matrix}
\bibfield{author}{\bibinfo{person}{G.W. Stewart} {and} \bibinfo{person}{J.G.
  Sun}.} \bibinfo{year}{1990}\natexlab{}.
\newblock \bibinfo{booktitle}{\emph{Matrix Perturbation Theory}}.
\newblock \bibinfo{publisher}{ACADEMIC PRESS, INC.}
\newblock
\showISBNx{9781493301997}


\bibitem[\protect\citeauthoryear{Tropp et~al\mbox{.}}{Tropp
  et~al\mbox{.}}{2015}]%
        {tropp2015introduction}
\bibfield{author}{\bibinfo{person}{Joel~A Tropp} {et~al\mbox{.}}}
  \bibinfo{year}{2015}\natexlab{}.
\newblock \showarticletitle{An introduction to matrix concentration
  inequalities}.
\newblock \bibinfo{journal}{\emph{Foundations and Trends{\textregistered} in
  Machine Learning}} \bibinfo{volume}{8}, \bibinfo{number}{1-2}
  (\bibinfo{year}{2015}), \bibinfo{pages}{1--230}.
\newblock


\bibitem[\protect\citeauthoryear{Varma and Yoshida}{Varma and Yoshida}{2019}]%
        {VY19:sensitivity}
\bibfield{author}{\bibinfo{person}{Nithin Varma} {and} \bibinfo{person}{Yuichi
  Yoshida}.} \bibinfo{year}{2019}\natexlab{}.
\newblock \showarticletitle{Average Sensitivity of Graph Algorithms}.
\newblock \bibinfo{journal}{\emph{CoRR}}  \bibinfo{volume}{abs/1904.03248}
  (\bibinfo{year}{2019}).
\newblock
\showeprint[arxiv]{1904.03248}


\bibitem[\protect\citeauthoryear{von Luxburg}{von Luxburg}{2007}]%
        {Luxburg2007}
\bibfield{author}{\bibinfo{person}{Ulrike von Luxburg}.}
  \bibinfo{year}{2007}\natexlab{}.
\newblock \showarticletitle{A tutorial on spectral clustering}.
\newblock \bibinfo{journal}{\emph{Statistics and Computing}}
  \bibinfo{volume}{17}, \bibinfo{number}{4} (\bibinfo{year}{2007}),
  \bibinfo{pages}{395--416}.
\newblock


\bibitem[\protect\citeauthoryear{Zhang and Rohe}{Zhang and Rohe}{2018}]%
        {zhang2018understanding}
\bibfield{author}{\bibinfo{person}{Yilin Zhang} {and} \bibinfo{person}{Karl
  Rohe}.} \bibinfo{year}{2018}\natexlab{}.
\newblock \showarticletitle{Understanding regularized spectral clustering via
  graph conductance}. In \bibinfo{booktitle}{\emph{NeurIPS}}.
  \bibinfo{pages}{10631--10640}.
\newblock


\end{thebibliography}

\appendix

\section{Missing Proofs of Section~\ref{sec:pre}}
\subsection{Proof Sketch of Lemma~\ref{lem:improvedCheeger}}\label{sec:proof-of-improved-cheeger}
We slightly modify the proof of Lemma~\ref{lem:improvedCheeger} for the normalized case~\cite{KLLOT:arxiv}, which is given as Theorem~1.2 in Section~3.1 in~\cite{KLLOT:arxiv}.

We now start with $\bm{v}_2$, the second eigenvector of the (unnormalized) Laplacian $L$, with corresponding second smallest eigenvalue $\lambda_2$.
By an analogous argument in the proof of Corollary 2.2 of~\cite{KLLOT:arxiv}, we can find a non-negative function $f\colon \{1,\ldots,n\} \to \mathbb{R}$ with Rayleigh quotient $R(f)\leq \lambda_2$ and $|\supp(f)|\leq n/2$, and $\norm{f}_2^2=1$, where we identify $f$ with a vector in $\mathbb{R}^n$ and the \emph{Rayleigh quotient} $R\colon \mathbb{R}^n \to \mathbb{R}$ of $G$ is defined as $R(\bm{x})=\frac{\bm{x}^\top L_G \bm{x}}{\bm{x}^\top \bm{x}}$.

Then we can find a $(2k+1)$-step function $g$ that well approximates $f$. By analogous argument in the proof of Lemma~3.1 in~\cite{KLLOT:arxiv}, we can find such a step function $g \colon \{1,2,\ldots,n\} \to \mathbb{R}$ such that
$\norm{f-g}_2^2\leq \frac{4 R(f)}{\lambda_k}$, where $\lambda_k$ is the $k$-th smallest eigenvalue of $L$.
(In our case, we need to consider $\norm{f-g}_2$, rather than the quantity $\norm{f-g}_w$ that takes the degree of each vertex into account as in~\cite{KLLOT:arxiv}: e.g., in inequality~(3.2) in~\cite{KLLOT:arxiv}, we do not need the factor $w(v)$.
Furthermore, we also need the fact that for any $k$ disjointly supported functions $f_1,\ldots,f_k$, we have $\lambda_k\leq 2 \max_{1\leq i\leq k} R(f_i)$, whose proof directly follows from the proof of Lemma~2.3 of~\cite{KLLOT:arxiv}.)

Then, we can also find a function $h\colon \{1,2,\ldots,n\}\to \mathbb{R}$ that defines the same sequence of threshold sets as $f$ (and thus we have $\alpha(\sweep_\alpha(h))=\alpha(\sweep_\alpha(f))$), and satisfies that
\[
  \frac{\sum_{u\sim v}|h(u)-h(v)|}{\sum_v h(v)}\leq 4k R(f) + 4\sqrt{2}k \sqrt{\Delta} \norm{f-g}_2\sqrt{R(f)}.
\]
This follows from an analogous argument for proving the Proposition 3.2 in~\cite{KLLOT:arxiv}. (The main difference is that, again, we use the $\norm{\cdot}_2$ norm rather than the $\norm{\cdot}_w$ norm, and after the third inequality of the proof of Proposition 3.2 on page 16 in~\cite{KLLOT:arxiv}, we will use the fact that $\sum_{u\sim v}(|f(u)-g(u)|^2+|(f(v)-g(v))|^2)\leq 2\Delta \norm{f-g}_2^2$.)

Then finally, we use the fact that for any non-negative function $h$ such that $|\supp(h)|\leq \frac{n}{2}$, it holds that
$\alpha(\sweep_\alpha(h))\leq \frac{\sum_{u\sim v}|h(u)-h(v)|}{\sum_v h(v)}$.
 (This follows directly from the proof of Lemma 2.4 in~\cite{KLLOT:arxiv}).
Then, we have $\alpha(\sweep_\alpha(f)) = \alpha(\sweep_\alpha(h))$, which is at most
  $4k \lambda_2 + 8\sqrt{2}k\sqrt{\Delta}\frac{\lambda_2}{\lambda_k}
    \leq 12\sqrt{2}k\lambda_2\sqrt{\frac{\Delta}{\lambda_k}}$.

\vspace{-1em}
\subsection{Proof of Lemma~\ref{lemma:stable_kllot_Delta}}\label{sec:proof-of-stable_kllot_Delta}

Consider an arbitrary optimal solution $T \subseteq V$ with $|T|\leq n/2$, and let $\alpha = \alpha(T)$.
Suppose that there exists a set $S \subseteq V$ with $|S|\leq n/2$ and $\alpha(S)\leq \rho\alpha$ satisfying that $d_{\mathrm{size}}(S,T) \geq \varepsilon $ for some $0<\varepsilon\leq 1/2$.
Let $S_1$ be either $S \setminus T$ or $T \setminus S$, whichever of a larger size.
Let $S_2$ be either $S\cap T$ or $V \setminus (S \cup T)$, whichever of a larger size.
Then by our assumption, we have that $n/2 \geq |S_1|\geq |S\triangle T|/2\geq \varepsilon n/2$, $|S_2|\geq (n-|S\triangle T|)/2\geq \varepsilon n/2$ and $|S_2|\leq \max\{|S\cap T|, |V \setminus (S\cup T)|\}\leq \max\{n/2,(1-\varepsilon)n \}=(1-\varepsilon)n
$.
Furthermore, for each $i=1,2$, $|E(S_i,V \setminus S_i)|\leq |E(S, V \setminus S)|+ |E(T, V \setminus T)|\leq \alpha |T| + \rho\alpha |S|\leq (1+\rho)\alpha n/2$.
Therefore, $\alpha(S_i)\leq \frac{(1+\rho)\alpha n/2}{\varepsilon n/2} \leq (1+\rho)\alpha/\varepsilon$.

Now let $S_3 = V \setminus (S_1 \cup S_2)$, which is one of the four sets $T,S,V\setminus T, V \setminus S$, and thus $\alpha(S_3)\leq \rho\alpha/\varepsilon$.
Therefore, $\lambda_3 \leq 2\max_i\alpha(S_i)\leq\frac{2\rho\alpha}{\varepsilon} \leq c_0\frac{\rho \lambda_2}{\varepsilon}\sqrt{\Delta/\lambda_k}$, for some constant $c_0> 0$.
Thus $\varepsilon \leq c_0 \rho \lambda_2 \sqrt{\Delta/\lambda_k^3}$.
This further implies that $G$ is $(\rho, \Theta(\rho \lambda_2 \sqrt{\Delta/\lambda_3^3}))$-stable.

\vspace{-0.5em}
\section{Proof of Theorem~\ref{thm:normalized_random_deletion_multiple}}\label{sec:proof-normalized_random_deletion_multiple}

Now we prove Theorem~\ref{thm:normalized_random_deletion_multiple}. Instead of proving that the statement of the theorem holds under Assumption~\ref{assume:normalized}, we prove it under a weaker assumption.

Let $G=(V,E)$ be a graph with minimum degree $\tau$ and $\Gamma:=\max_{i\in V}\sum_{j:(j,i)\in E}d_j^{-2}$. Let
\begin{align*}
q := \Theta\left(\max\left\{\sqrt{p\left(\frac{1}{\tau}+\Gamma\right)\log n}, \frac{\log n}{\tau}\right\}\right)\text{ and }
p_\diamond :=p + q.
\end{align*}
We need the following assumption: %
\begin{assumption}\label{assume:normalized_weak}
	\begin{itemize}
		\item[] $\nu_3(G)\geq 6p_\diamond$.
	\end{itemize}
\end{assumption}

Note that any graph $G$ satisfying Assumption~\ref{assume:normalized}(i) and (iii) also satisfies Assumption~\ref{assume:normalized_weak}. This is true as $\nu_3(G) \geq \Omega(\tau^{-1}\log n)$, $\Gamma\leq  \frac{\Delta}{\tau^2}$ and $p=O(\Delta^{-1}\log n)$, which gives that
\[
p_\diamond = O\left(\max\left\{\sqrt{p\left(\frac{1}{\tau}+\Gamma\right)\log n}, \frac{\log n}{\tau}\right\}\right)\leq O\left(\frac{\log n}{\tau}\right),
\]
and thus that $\nu_3(G) \geq 6p_\diamond$. Therefore, Theorem~\ref{thm:normalized_random_deletion_multiple} follows from the following theorem, which we prove in the rest of this section.

\begin{theorem}\label{thm:normalized_random_deletion_multiple_weak}
	Let $G=(V,E)$ be a graph and $F \sim_p E$.
	If Assumption~\ref{assume:normalized_weak} holds, then we have $\nu_3(G-F)\geq \nu_3(G)/2$ with probability at least $1-n^{-7}$.
\end{theorem}

We first introduce some definitions. For each $e=(i,j)\in E$, we let $E_{1,e}=E_{ii}+E_{jj}$ and $E_{2,e}=E_{ij}+E_{ji}$. Note that $\sum_{e\in E}E_{1,e}=D$ and $\sum_{e\in E}E_{2,e}=A$.
For a set of edges $F$, let $E_{1,F}=\sum_{e=(i,j)\in F}E_{1,e}$, $E_{2,F}=\sum_{e\in F}E_{2,e}$. Note that $E_F=E_{1,F}-E_{2,F}$ is the unnormalized Laplacian of a graph $(V,F)$.
We will make use of the following matrix Chernoff bound.
\begin{theorem}[Corollary 6.1.2 in~\cite{tropp2015introduction}]\label{thm:matrix_ineq}
  Let $X = \sum_{i=1}^T X_i$, where $X_i \in \mathbb{R}^{n \times n}\;(1\leq i \leq T)$ are independent random matrices. Assume that $\norm{X_i-\E[X_i]}\leq L$ holds for every $1 \leq i \leq T$.
  Let $Y_i=X_i-\E[X_i]$ and $\nu(X)  = \max\left\{\norm{\sum_i\E[Y_i Y_i^\top]}, \norm{\sum_i\E[Y_i^\top Y_i]}\right\}$.
  Then for any $t > 0$, we have
  \[
    \Pr[\norm{X-\E[X]}\geq t] \leq 2n\cdot \exp\left(-\frac{t^2/2}{\nu(X)+Lt/3}\right).
  \]
\end{theorem}

Now we give some claims.
\begin{claim}
  With probability at least $1- n^{-8}$,
  \begin{eqnarray}
    \norm{D^{-1}E_{2,F}} \leq p_\diamond.\label{eqn:norm_bound_1}
  \end{eqnarray}
\end{claim}
\begin{proof}
  For any $e\in E$, let $X_e$ be the indicator random variable of the event that $e$ is included in $F$.
  Note that \[E_{2,F}=\sum_{e\in F}E_{2,e}=\sum_{e\in E}X_e \cdot E_{2,e} \Longrightarrow D^{-1}E_{2,F}=\sum_{e\in E}X_e \cdot D^{-1}E_{2,e}.\]
  Then by $\Pr[X_e=1]=p$, we have $\E[D^{-1}E_{2,F}]=pD^{-1}\sum_{e\in E}E_{2,e}=pD^{-1}A$.

  Now note that the variables $X_e\;(e\in E)$ are independent and thus $D^{-1}E_{2,F}$ is a sum of independent random variables $S_e:=X_e \cdot D^{-1}E_{2,e}$.
  We further note that $\E[S_e]=pD^{-1}E_{2,e}$.
  \[
  \norm{S_e-\E[S_e]}=\norm{\left(X_e-p\right)D^{-1}E_{2,e}}\leq \max\{1-p,p\}\cdot\norm{D^{-1}E_{2,e}}\leq \frac{1}{\tau},
  \]
  as we assume that $\min_v\deg(v)\geq \tau$.

  Now we note that $E_{2,e}^2=E_{1,e}$ and thus
  \begin{align*}
    & (S_e-\E[S_e]){(S_e-\E[S_e])}^\top = {\left(X_e-p\right)}^2D^{-1}E_{2,e}^2 D^{-1} \\
    & ={\left(X_e-p\right)}^2 D^{-1}E_{1,e}D^{-1} ={\left(X_e-p\right)}^2 D^{-2}E_{1,e}.
  \end{align*}
  Thus
  \[
    \E[(S_e-\E[S_e]){\left(S_e-\E[S_e]\right)}^\top]=\left(p{\left(1-p\right)}^2+p^2\left(1-p\right)\right)D^{-2}E_{1,e}.
  \]
  Recall that $\sum_e E_{1,e}=D$. We have that
  \begin{align*}
    & \norm{\sum_e\E[(S_e-\E[S_e]){(S_e-\E[S_e])}^\top]} \\
    & = \left(p{\left(1-p\right)}^2+p^2\left(1-p\right)\right)\norm{\sum_e D^{-2}E_{1,e}}\leq p\norm{D^{-1}}\leq \frac{p}{\tau}.
  \end{align*}

  Similarly, ${(S_e-\E[S_e])}^\top (S_e-\E[S_e]) = {\left(X_e-p\right)}^2E_{2,e} D^{-2}E_{2,e}$ holds. %
  Recall that $\Gamma=\max_{i\in V}\sum_{j:(j,i)\in E}\frac{1}{d_j^2}$. Note that $\sum_e E_{2,e}D^{-2}E_{2,e}$ is a diagonal matrix such that each diagonal entry is at most $\Gamma$. %
Then 
  \begin{align*}
    & \norm{\sum_e\E[{(S_e-\E[S_e])}^\top (S_e-\E[S_e])]} \\
    & = \left(p{\left(1-p\right)}^2+p^2\left(1-p\right)\right)\norm{\sum_e E_{2,e}D^{-2}E_{2,e}}\leq p \Gamma. %
  \end{align*}

By the matrix Chernoff bound (Theorem~\ref{thm:matrix_ineq}) with $X_i=S_e$, $X=D^{-1}E_{2,F}$, $\nu(X)=p\max\{1/\tau,\Gamma\}$, $L=1/\tau$, we have that
  for any $t>0$,
  \begin{align*}
    & \Pr\left[\norm{D^{-1}E_{2,F}-pD^{-1}A}\geq t\right] \leq 2n\cdot \exp\left(-\frac{t^2/2}{\nu(X)+Lt/3}\right) \\
    & \leq 2n\exp\left(-\frac{t^2/2}{p\max\{\frac{1}{\tau},\Gamma\}+\frac{t}{3\tau}}\right).
  \end{align*}
  By setting $t= q$, we have that with probability at least $1-n^{-8}$
  \begin{eqnarray*}
  \norm{D^{-1}E_{2,F}-pD^{-1}A} \leq q.
  \end{eqnarray*}
  Thus,
  \begin{align*}
  & \norm{D^{-1}E_{2,F}}\leq p\norm{D^{-1}A}+q \leq p+q = p_\diamond. \qedhere
  \end{align*}
\end{proof}

\begin{claim}
  With probability at least $1-n^{-8}$
  \begin{eqnarray}
    \norm{D^{-1}E_{1,F}} \leq p_\diamond \label{eqn:norm_bound_2}
  \end{eqnarray}
\end{claim}
\begin{proof}
The proof is similar to the above. Note that
\[D^{-1}E_{1,F}=\sum_{e\in F}D^{-1}E_{1,e} =\sum_{e\in E}X_e \cdot D^{-1}E_{1,e}
\]
That is, since the variables $X_e\;(e\in E)$ are independent, $D^{-1}E_{1,F}$ is a sum of independent random variables $S_e:=X_e \cdot D^{-1}E_{1,e}$. Note that $\E[S_e]=p \cdot D^{-1}E_{1,e}$. We further note that
\[
\E[D^{-1}E_{1,F}] =p \sum_{e\in E} D^{-1}E_{1,e} =p \cdot D^{-1}D=p
\]

Then by similar calculations to the proof of the previous claim, we bound $\norm{S_e-\E[S_e]} \leq \frac{1}{\tau}$ and further by the fact that $E_{1,e}^2=E_{1,e}$, we have
\begin{align*}
& \norm{\sum_e\E[{(S_e-\E[S_e])}{(S_e-\E[S_e])}^\top ]}\leq \frac{p}{\tau}.
\end{align*}
Now since $E_{1,e},E_{1,F},D^{-1}$ are all diagonal matrices, we have
\begin{align*}
& \sum_e\E[{(S_e-\E[S_e])}^\top{(S_e-\E[S_e])}]\\
&= \left(p{\left(1-p\right)}^2+p^2\left(1-p\right)\right)\sum_e E_{1,e} D^{-2}E_{1,e}\\
&=\left(p{\left(1-p\right)}^2+p^2\left(1-p\right)\right)\sum_e E_{1,e} D^{-2}
\end{align*}
Thus by similar analysis as before, we have
\begin{align*}
\norm{\sum_e\E[{(S_e-\E[S_e])}^\top(S_e-\E[S_e]) ]}\leq \frac{p}{\tau}.
\end{align*}

Then the rest follows the same as before.
\end{proof}

\vspace{-1em}
\begin{proof}[Proof of Theorem~\ref{thm:normalized_random_deletion_multiple_weak}]
We set $p_\diamond$ to be the maximum of the RHS of Ineq. (\ref{eqn:norm_bound_1}) and the RHS of Ineq. (\ref{eqn:norm_bound_2}). Then by the above two claims, we have $\norm{D^{-1}E_{2,F}}\leq p_\diamond$ and $\norm{D^{-1}E_{1,F}}\leq p_\diamond$ with probability at least $1-2 \cdot n^{-8}$.
We will condition on the above two inequalities in the following. %

We have
\begin{align}
& \mathcal{L}_{G-F} %
= I- D_{G-F}^{-1}A_{G-F}=I-{(D-E_{1,F})}^{-1}(A-E_{2,F}) \nonumber \\
&=I-{(I-D^{-1}E_{1,F})}^{-1}D^{-1}(A-E_{2,F})\nonumber\\
&=
I-\sum_{i\geq 0}{(D^{-1}E_{1,F})}^i\cdot D^{-1}(A-E_{2,F}) \tag{by $\norm{D^{-1}E_{1,F}}\leq p_{\diamond}<1$}\\
&= I-\left(I+\sum_{i\geq 1}{(D^{-1}E_{1,F})}^i\right)D^{-1}(A-E_{2,F}) \nonumber %
\\
&=I-\left(D^{-1}+\sum_{i\geq 1}{(D^{-{1}}E_{1,F})}^i D^{-1}\right)(A-E_{2,F})\nonumber\\
&=I-D^{-1}A+D^{-1}E_{2,F}-\sum_{i\geq 1}{(D^{-{1}}E_{1,F})}^i\cdot D^{-1}(A-E_{2,F})\nonumber\\
&\stackrel{}{=}\mathcal{L}+D^{-1}E_{2,F}-\sum_{i\geq 1}{(D^{-{1}}E_{1,F})}^i\cdot D^{-1}(A-E_{2,F})\nonumber %
\end{align}
Now, we have that
\begin{align*}
  & \norm{D^{-1}E_{2,F}-\sum_{i\geq 1}{(D^{-{1}}E_{1,F})}^i \cdot D^{-1}(A-E_{2,F})} \\
  &\leq \norm{D^{-1}E_{2,F}} + \sum_{i\geq 1} \norm{D^{-{1}}E_{1,F}}^{i}\norm{D^{-1}(A-E_{2,F})}\\
  &\leq p_\diamond + \sum_{i\geq 1} \norm{D^{-{1}}E_{1,F}}^{i}\leq p_\diamond+\sum_{i\geq 1}p_\diamond^i\leq p_\diamond+\frac{3}{2}p_\diamond<3p_\diamond,
\end{align*}
where the penultimate inequality follows from the assumption that $p_\diamond\leq \frac16\nu_3(G)\leq \frac{1}{3}$.

Finally, by Weyl's inequality, %
\begin{align*}
  & \nu_3(G-F) \geq \nu_3(G)- \norm{D^{-1}E_{2,F}-\sum_{i\geq 1}{(D^{-{1}}E_{1,F})}^i \cdot D^{-1}(A-E_{2,F})} \\
  & > \nu_3(G)- 3p_\diamond\geq \frac{\nu_3(G)}{2},
\end{align*}
where the last inequality follows from Assumption~\ref{assume:normalized_weak}.\qedhere
\end{proof}

\section{Missing Proofs of Section \ref{sec:kcluster}}\label{sec:proof-new_stable_kllot_Delta}
Now we sketch the proof of Lemma \ref{lemma:new_stable_kllot_Delta}.
\vspace{-0.5em}
\begin{proof}[Proof Sketch of Lemma \ref{lemma:new_stable_kllot_Delta}]
The proof follows by adapting the proof of Lemma \ref{lemma:stable_kllot_Delta}, i.e., Corollary 4.17 in~\cite{KLLOT:arxiv}, which considers the case $k=2$. For $k>2$, we only need to start with an optimum solution $\mathcal{S}^*=\{S_1^*,\ldots,S_k^*\}$ with $\rho_G(\mathcal{S})=\rho_G(k)$. Then we assume the instance is not $(c,\varepsilon)$-stable, and then construct as in~\cite{KLLOT:arxiv} a $(k+1)$-partition $T_1,\ldots,T_{k+1}$ such that $\nu_{k+1}(G) \leq 2\max_{i}\phi(T_i)=O(ck^3\rho_G(k)/\varepsilon)$. Then we conclude that $\varepsilon=O(ck^3\rho_G(k)/\nu_{k+1}(G))$.
\end{proof}

\end{document}